\documentclass[12pt]{article}
\usepackage[utf8]{inputenc}
\usepackage[margin=0.8in]{geometry}
\usepackage{blindtext, xfrac}
\usepackage[english]{babel}
\usepackage[T1]{fontenc}
\usepackage{titling}
\usepackage{amsmath}
\usepackage{amssymb}
\usepackage{cite}
\usepackage{tikz}
\usepackage{graphicx}
\usepackage{booktabs}
\usepackage{physics}
\usepackage{mathtools}
\usepackage{nameref}
\usepackage{bbold}
\usepackage{textgreek}
\usepackage{nicematrix}
\usepackage{hyperref}
\usepackage{gensymb}
\usepackage{float}
\usepackage{svg}
\usepackage{bm}
\usepackage{placeins}
\usepackage{tikz-feynman}
\usetikzlibrary{decorations.pathreplacing,decorations.markings}

\definecolor{greeny}{rgb}{0,0,1}

\newcommand*\diff{\mathop{}\!\mathrm{d}}
\newcommand*\Diff[1]{\mathop{}\!\mathrm{d^#1}}
\DeclareMathOperator{\arctanh}{arctanh}
\DeclareMathOperator{\arccoth}{arccoth}
\DeclareMathOperator{\Log}{Log}

\DeclareRobustCommand{\stirling}{\genfrac\{\}{0pt}{}}

\title{Smooth Perturbations to Rényi Entropy}

\author{ Andrew Buchanan$^{1,2}$
\\
$^1$\small Department of Physics, Engineering Physics, \& Astronomy \protect\\ 
\small Queen's University, Kingston, ON, K7L 2S8, Canada
\\
$^2$\small Arthur B. McDonald Canadian Astroparticle Physics Research Institute, \protect\\ 
\small Kingston, ON, K7L 3N6, Canada
}
\begin{document}
\maketitle

\begin{abstract}

A method is presented for computing the Rényi entropy of a perturbed massless vacuum on the ball via a comparison with lattice field theory. If the perturbed state is Gaussian with smoothly varying correlation functions and the perturbation parameter has units of energy, I show the coefficients for Rényi entropy are analytically computable for all Rényi parameter $\alpha$ in odd dimensions and for integer $\alpha$ in even dimensions. I apply this procedure to compute coefficients for the large distant expansion for the Rényi mutual information of distant balls and the low temperature expansion for the entropy of a thermal field.

\end{abstract}

\tableofcontents

\section{Introduction}
In recent decades, there has been interest in understanding entanglement entropy in quantum scalar fields, motivated by an influential paper by Srednicki identifying a possible connection between it and quantum gravity. \cite{Srednicki_1993} The entanglement entropy, Von Neumann entropy or simply entropy of a quantum state $\hat\rho$ is 
\begin{equation}
    S(\hat\rho)  = -\Trace(\hat{\rho}\ln(\hat{\rho}))).
\end{equation}
The entanglement entropy is a special case of a larger family of entanglement measures collectively called the Rényi entropy, defined in terms of $\rho$ and a parameter $\alpha\in [0,\infty]$ by
\begin{equation}
    S_\alpha(\hat\rho) = \frac{1}{1-\alpha}\ln(\Trace(\hat{\rho}^\alpha)).
\end{equation}
Taking the limit as $\alpha$ tends to $1$ recovers the Von Neumann entropy. A related quantity is the mutual information on a joint entangled state on subsystems $1$ and $2$. If $\hat{\rho}_{12}$ is the full state and if $\hat{\rho}_{1}$ and $\hat{\rho}_{2}$ are its partial traces to each subsystem, the mutual information is given by 
\begin{equation}
   I(\hat{\rho}_{12}) = S(\hat{\rho}_1)+ S(\hat{\rho}_2) - S(\hat{\rho}_{12}).
\end{equation}
By replacing entropy with Rényi entropy in the above equation, we get the Rényi mutual information of which the standard mutual information is a special case.

Towards the aim of understanding entanglement in a scalar field, one problem which has received attention is quantifying the behavior of mutual information of distant balls. Specifically, given two distant balls of radius $r$ separated by distance $R$, the problem is to compute the highest order coefficients of the series,
\begin{equation}
   I(r,R)= \sum_{N=1}^\infty I_{d,\alpha}^{[N]}\left(\frac{r}{R}\right)^{N}.
\end{equation} 
While there has been some numerical work on this problem by Shiba \cite{Shiba_2012}, this is the rare problem which can be solved analytically. A strategy for computing these coefficients was derived in \cite{Cardy_2013} by using a replica trick to find the coefficients of Rényi mutual information for integer $\alpha\geq 2$, then analytically continuing to $\alpha=1$. Results for the leading coefficient can be found in \cite{Cardy_2013,Agon2015,Chen_2018} and results for next to leading coefficients can be found in \cite{Chen_2017,chen_odd,Ag_n_2016}.  However, in \cite{firstPaper}, a coauthor and I introduced a novel strategy to compute this series which was inspired by the numerical approaches to approximate entropy using lattice field theory. We showed that this problem can be reframed as an example of the following problem:
\begin{flushleft}
    Given a smooth family of Gaussian states on the ball $\hat\rho(t)$, where $t$ has units of energy and $\hat\rho(0)$ is the ground state of a massless scalar field on a ball of radius $r$, find the leading order terms of the expansion for $S(\hat\rho(t))-S(\hat\rho(0))$ in terms of $ rt$.
\end{flushleft}
A Gaussian state is a special class of state completely described by its two-point correlation functions of the field $\hat\phi$ and its momentum conjugate $\hat\pi$. In a lattice field theory with $N$ sites, the two point correlation functions are convenient to combine into a single $2N\times 2N$ \textit{correlation matrix}, denoted $\sigma$, whose entries are, 
\begin{equation}\label{eq:latticeCorrMatrix}
\sigma_{ij} =
\begin{cases} 
  2\langle \hat{\phi}_i\hat{\phi}_j\rangle & \text{if } i,j\leq N, \\
2\langle \hat{\pi}_{i-N}\hat{\pi}_{j-N}\rangle & \text{if } i,j>N, \\
\langle\{\hat{\phi}_j, \hat{\pi}_{i-N} \}\rangle & \text{if } i>N \text{ and } j\leq N,\\
\langle\{\hat{\phi}_i, \hat{\pi}_{j-N} \}\rangle & \text{if } i\leq N \text{ and } j>N,\\
\end{cases}
\end{equation}
where $\{\cdot,\cdot\}$ is the anti-commutator and $\langle\cdot\rangle$ is the expectation value over the state.
The entanglement entropy on a lattice can be written as the following trace of a matrix function involving the correlation matrix.
\begin{equation}\label{eq:matFun}
  S_\alpha(\hat{\rho})= \Trace\left[h(J_{\mathrm{lat}}\sigma)\right]
\end{equation}
where $h(z)=\frac{z}{2}\arccot(z)+\frac{1}{4}\ln(-\frac{1}{4}(z^2+1))$ and $J_{\mathrm{lat}}=
    \begin{bmatrix}
     0 &I \\-I&0
 \end{bmatrix}$. In \cite{firstPaper}, we generalized this formula to the continuum limit. In this limit, the correlation matrix becomes a linear map on the vector space of 2-vector fields on the ball and $J_{\mathrm{lat}}$ becomes the linear map defined by \begin{equation}\label{eq:J}
J\begin{bmatrix}\phi_1\\\phi_2\end{bmatrix} =\begin{bmatrix}
    \phi_2\\-\phi_1
\end{bmatrix}.
\end{equation} 
The correlation matrix $\sigma$ is easiest to represent with a $2\times 2$ matrix kernel whose entries are the two point correltion functions,
\begin{equation}\label{eq:2x2kernel}
    K(\vectorbold{x},\vectorbold{y}) = 
    \begin{bmatrix}
        2\langle \hat{\phi}(\vectorbold{x})\hat{\phi}(\vectorbold{y})\rangle &  \langle\{\hat{\phi}(\vectorbold{x}),\hat{\pi}(\vectorbold{y})\}\rangle \\\langle\{\hat{\phi}(\vectorbold{y}),\hat{\pi}(\vectorbold{x})\}\rangle & 2\langle  \hat{\pi}(\vectorbold{x})\hat{\pi}(\vectorbold{y})\rangle \\
    \end{bmatrix}.
\end{equation}
where the corresponding linear map $\sigma$ is
\begin{equation}
     \sigma\begin{bmatrix}\phi_1\\\phi_2\end{bmatrix}(\vectorbold{y}) =\int_\Omega \diff^d \vectorbold{x} \text{ } K(\vectorbold{x},\vectorbold{y})\begin{bmatrix}\phi_1(\vectorbold{x})\\\phi_2(\vectorbold{x})\end{bmatrix},
\end{equation}
where $\Omega$ is the ball of radius $r$.
These identifications allowed us to construct a perturbative series for the entropy in terms of the kernels of the correlation matrices.
In \cite{firstPaper}, we apply this procedure to compute the lowest order term of the mutual information of distant balls, setting $t=R^{-1}$. We also treated a new problem. A massless scalar field at temperature $T$ is the Gaussian state whose correlation matrix has kernel,

\begin{equation}\label{eq:themalCorrelators}
     K(\vectorbold{x},\vectorbold{y};T)=\frac{1}{(2\pi)^d}\int_{\mathbb{R}^d} \Diff{d} \vectorbold{k} \text{ }e^{i\vectorbold{k}\cdot(\vectorbold{x}-\vectorbold{y})}\begin{bmatrix}
        \frac{1}{|\vectorbold{k}|}\coth(\frac{ |\vectorbold{k}|}{2T}) &0 
   \\0 &|\vectorbold{k}|\coth(\frac{ |\vectorbold{k}|}{2T})
    \end{bmatrix}\footnote{Compare this with the expectation values of $\hat{X}^2$ and $\hat{P}^2$ for the canonical ensemble of a one dimensional harmonic oscillator at temperature $T$.}.
\end{equation}
In \cite{firstPaper}, we found the first three leading order terms of the entropy difference $S(\hat\rho(T))-S(\hat\rho(0))$ in terms of $ rT$ by noting that $T=0$ corresponds to the vacuum state.

In this work, I present a refinement of the procedure outlined in \cite{firstPaper} which provided a strategy for computing the lowest order term of the expansion which does not generalize easily to higher order terms. I show that, under an additional smoothness assumption, for a fixed dimension, arbitrarily many terms of the expansion for Von Neumann entropy can be computed. Under this additional assumption, computation of the coefficients is reduced to a finite dimensional linear algebra problem and a few countable families of integrals, all of which can be computed analytically.
I also generalize to the case of Rényi entropy and mutual information, and show that the terms of a series can be computed exactly for all $\alpha$ in odd spatial dimensions and for integer $\alpha$ and $\alpha=\infty$ for even spatial dimensions. I then apply this result to compute three nonzero terms in the series for the Rényi mutual information of distant identical balls and five nonzero terms for the entropy difference of a thermal field, which is two more nonzero terms than the results presented in \cite{firstPaper}.

The layout of this work is as follows. Section \ref{sec:entropyExpans} details the general procedure to compute the series of an entropy difference. Sections \ref{sec:MI} examine the Rényi mutual information of distant balls. The leading order coefficients of standard mutual information ($\alpha=1$) are presented for spatial dimensions $2\leq d\leq 6$. Section \ref{sec:Ent} does the same for the Rényi entropy of a thermal field on a ball and presents the coefficients for the Von Neumann entropy for spatial dimensions $3\leq d \leq 6$. Section \ref{sec:conlusion} presents general conclusions and potential future directions. Appendix \ref{sec:renyi} calculates the generalization of Equation \ref{eq:matFun} to arbitrary Rényi parameter $\alpha$, which allows the continuum limit to be taken in Section \ref{sec:entropyExpans}. Appendix \ref{polybasis} describes a convenient multivariate polynomial basis used in this work. Appendix \ref{sec:harmPolys} provides formulae for the Cartesian form of low dimensional spherical harmonics. Computation of the integrals needed in this work is in Appendix \ref{sec:integrals}. Appendix \ref{sec:bernoulli} covers the Bernoulli polynomials and Bernoulli numbers which appear in these integrals. Lastly, Appendix \ref{sec:RenyiFull} provides the results of Sections \ref{sec:MI} and \ref{sec:Ent} for Renyi parameter $\alpha \neq 1$.

\section{Computing the Terms of a Rényi Entropy Expansion} \label{sec:entropyExpans}
In \cite{firstPaper}, we considered a family of Gaussian states on the ball $\hat\rho(t)$, where $t$ has units of energy and $\hat\rho(0)$ is the ground state of a massless scalar field on a ball of radius $r$.
  In this work, I make the additional assumption that the two-point correlation functions are smoothly varying. More precisely, assume that the difference in correlation functions, 
\begin{align}\label{eq:sigComponents}
    \begin{split}
        \delta  X(\vectorbold{x},\vectorbold{y};t)&=\langle \hat{\phi}(\vectorbold{x})\hat{\phi}(\vectorbold{y})\rangle_{\hat\rho(t)}-\langle\hat{\phi}(\vectorbold{x})\hat{\phi}(\vectorbold{y})\rangle_{\hat\rho(0)}\\
       \delta  P(\vectorbold{x},\vectorbold{y};t) &=\langle\hat{\pi}(\vectorbold{x})\hat{\pi}(\vectorbold{y})\rangle_{\hat\rho(t)}-\langle\hat{\pi}(\vectorbold{x})\hat{\pi}(\vectorbold{y})\rangle_{\hat\rho(0)}\\
        \delta  V_{off}(\vectorbold{x},\vectorbold{y};t)&=\langle\{\hat{\phi}(\vectorbold{x}),\hat{\pi}(\vectorbold{y})\}\rangle_{\hat\rho(t)}-\langle\{\hat{\phi}(\vectorbold{x}),\hat{\pi}(\vectorbold{y})\}\rangle_{\hat\rho(0)},
    \end{split}
\end{align}
is analytic in $\vectorbold{x}$, $\vectorbold{y}$ and $\vectorbold{t}$ at zero. This is true for both examples I examine in Section \ref{sec:MI} and \ref{sec:Ent}. My goal is to compute the coefficients of the Taylor series of the Rényi entropy \begin{equation}\label{contExpan}
    S_\alpha(\hat\rho(t))-S_\alpha(\hat\rho(0)) = \sum_{N=1}^\infty S_\alpha^{[N]}(rt)^N.
\end{equation} 
Note that by the scale invariance of the theory, we can set the units so that $r=1$ for ease of calculation. I will do so for the remainder of this work, only reinserting units when presenting the final results.
The smoothness assumption on the correlation functions allow us to expand the correlation matrix of the $\hat\rho(t)$ as \begin{equation}\label{eq:sigExpa}
   \delta\sigma \coloneq \sigma(t) - \sigma_0=\sum_{k=1}^\infty \sigma^{(k)} t^k.
\end{equation} For brevity, $\sigma_0$ denotes the skewed correlation matrix of the vacuum on the unit ball. There is a corresponding expansion in the $2\times 2$ matrix kernel of $ \delta\sigma$, \begin{equation}\label{eq:kDecomp}
     K(\vectorbold{x},\vectorbold{y};t)=\sum_{k=1}^\infty K^{(k)}(\vectorbold{x},\vectorbold{y}) t^k.
 \end{equation}

  \begin{figure}[htbp]
    \centering
    \begin{tikzpicture}[scale=1.2]
        \draw[ ->] (-3.2,0) -- (3.2,0) node[right] {$\Re$};
        \draw[ ->] (0,-3.2) -- (0,3.2) node[above] {$\Im$};
        
        \draw[thick,->] (-0.15, 1)-- (-0.15, 3);
        \draw[thick,<-] (0.15, 1)--(0.15, 3);
        \draw[thick,->] (-0.15, -1)-- (-0.15, -3);
        \draw[thick,<-] (0.15, -1)-- (0.15, -3);

        \draw[line width=0.8mm, dashed, red] (0,-1) -- (0,-3);
        \draw[line width=0.8mm, dashed, red] (0,1) -- (0,3);
        \draw[line width=0.8mm, dashed, blue] (0,-1) -- (0,1);
        
         \node at (0,1)[circle,fill,inner sep=2pt]{};
         \node at (0,-1)[circle,fill,inner sep=2pt]{};
    \end{tikzpicture}
    \caption{A disconnected contour  close to the portion of the imaginary line with magnitude greater than one. I am focused on the limiting case when the lines approach the portion of the imaginary axis with a red dotted line. Integration is taken in the direction of the arrows. The red dotted line represents the spectrum of $J\sigma_0$. The blue dotted line represents the branch cut of $h'_\alpha$ in the general case.}
    \label{fig:double_keyhole}
\end{figure}
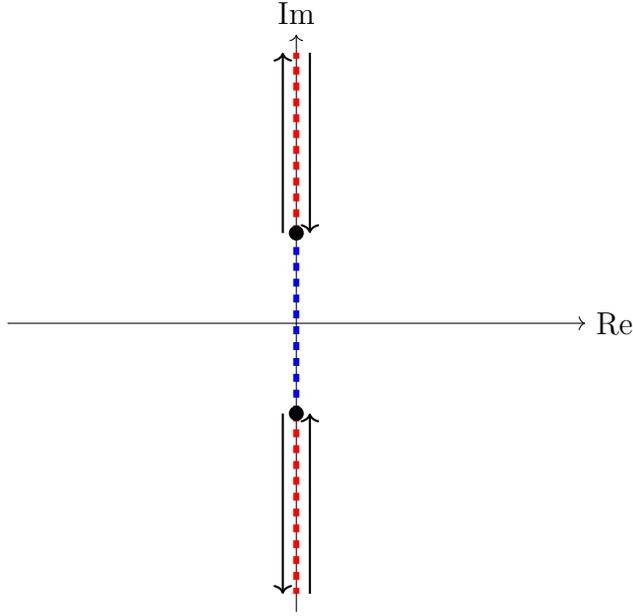

\subsection{Computing the Expansion for Rényi Entropy via Lattice Field Theory}
Consider a lattice field theory approximation for our problem, with lattice spacing $\epsilon$ and $D_\epsilon$ lattice points. Given the family of correlation matrices $\sigma(t)$ and the coefficients $\sigma^{(k)}$, let $\sigma_{\epsilon}(t)$ and $\sigma^{(k)}_\epsilon$ be their finite dimensional matrix approximations on the lattice. To find an expression for the $S_\alpha^{[N]}$, I mirror the strategy of \cite{firstPaper} by finding an approximate expression $S_{a,\epsilon}^{[N]}$ in lattice field theory in terms of the $\sigma_{\epsilon}$ and then taking the continuum limit. 

In Appendix \ref{sec:renyi}, I derived the following approximation for the Rényi entropy on a lattice
\begin{equation}\label{eq:contourEntropy}
     S_{\alpha,\epsilon}(t) = \frac{1}{2\pi i }\oint_\xi \diff z \text{ }h_\alpha(z)\Tr((z-J_{D_\epsilon}\sigma_\epsilon(t))^{-1}).
\end{equation}
 Here, $\xi$ is the non-closed contour shown in Figure \ref{fig:double_keyhole}, $J_{D_\epsilon}= \begin{bmatrix}
     0 &I_{D_\epsilon} \\-I_{D_\epsilon}&0
 \end{bmatrix} $  and $h_\alpha$ is the holomorphic function:  
\begin{equation}\label{eq:h_alpha}
   h_\alpha(z) = \frac{1}{4(\alpha - 1)} \Log \left( \left(1 - \left( \frac{z - i}{z + i} \right)^\alpha \right)^2 \right) + \frac{\alpha}{4(\alpha - 1)} \Log \left( -\frac{1}{4} (z + i)^2 \right).
\end{equation} Following the strategy of \cite{firstPaper}, we can use Taylor's theorem and integration by parts to derive,
 \begin{equation}\label{eq:GeneralDecompisitionLattice}
         S_{\alpha,\epsilon}^{[N]} = \sum_{\substack{N=\sum_{j=1}^nk_j \\n\geq 1,\text{ } k_j\geq 1}}\frac{1}{2\pi n i }\oint_\xi \diff z \text{ }h_\alpha'(z)\Trace \left[\prod_{q=1}^n (z-J_{D_\epsilon}\sigma_{\epsilon,0})^{-1}J_{D_\epsilon}\sigma_\epsilon^{(k_q)} \right].
 \end{equation}But this generalizes very naturally to the continuum limit $\epsilon\to 0$:
 \begin{equation}\label{eq:GeneralDecompisition}
         S_\alpha^{[N]} = \sum_{\substack{N=\sum_{j=1}^nk_j \\n\geq 1,\text{ } k_j\geq 1}}\frac{1}{2\pi n i }\oint_\xi \diff z \text{ }h_\alpha'(z)\Trace \left[\prod_{q=1}^n (z-J\sigma_0)^{-1}J\sigma^{(k_q)} \right].
 \end{equation}
 Here $h_\alpha'$ is the holomorphic function 
\begin{equation}\label{eq:gaderivative}
h_\alpha'(z) = \frac{\alpha}{2(\alpha-1)}\left[\frac{e^{2i(\alpha-1)\arccot(z)}-1}{(z+i)e^{2i(\alpha-1)\arccot(z)}-(z-i)}\right],
\end{equation}
where $\arccot(z)$ is on its principal branch and $J$ is the linear map defined in Equation \ref{eq:J}.
This can be recast more straightforwardly as,
\begin{equation}\label{eq:GeneralDecomposition_n}
    S_{\alpha}(t) = \sum_{n=1}^\infty\frac{1}{2\pi n i }\oint_\xi \diff z \text{ }h_\alpha'(z)\Trace \left[\left((z-J\sigma_0)^{-1}J\delta\sigma\right)^n \right],
\end{equation}
where substituting in Equation \ref{eq:sigExpa} recovers Equation \ref{eq:GeneralDecompisition}.
\subsection{Computing the Expansion in the Continuum Limit}

 In \cite{firstPaper}, we gave some heuristic arguments describing how the traces appearing in Equation \ref{eq:GeneralDecompisition} should be interpreted in the continuum limit in the general case. However, assuming the analyticity of $K(\vectorbold{x},\vectorbold{y},t)$, the interpretation is straightforward. We can use unit analysis to deduce that the entries of each $K^{(k)}$ are homogeneous polynomials of $\vectorbold{x}$ and $\vectorbold{y}$. In particular, the $X$ entry has degree $k-d+1$, the $P$ entry has degree $k-d-1$, and the $V_{off}$ entries have degree $k-d$. Therefore, the $\sigma^{(k)}$ have finite rank. This means that the linear map $\prod_{q=1}^n (z-J\sigma_0)^{-1}J\sigma^{(k_q)}$ is also finite rank, meaning its trace is well defined.

Define $\mathbb{C}[x_1,\dots,x_d]$ to be the vector space of $d$-variate complex polynomials and denote by $\mathbb{C}[x_1,\dots,x_d]^2$ the Cartesian product of this space with itself. Since each $\sigma^{(k)}$ is a polynomial, we can treat them as operators on $\mathbb{C}[x_1,\dots,x_d]^2$. To compute the trace of $\prod_{q=1}^n (z-J\sigma_0)^{-1}J\sigma^{(k_q)}$, we can expand each $\sigma^{(k)}$ using a basis for $\mathbb{C}[x_1,\dots,x_d]^2$. While we could use any basis including the standard monomial basis, it will be particularly convenient to expand in the basis described in Appendix \ref{polybasis}. This basis induces a decomposition of $\sigma^{(k)}$ into a finite sum,
\begin{equation}\label{sigmaDecomposition}
   \sigma^{(k)} = \sum_{\substack{\eta,\eta',j,j'\\|\ell|+|\ell'|+j+j' =k-d+1}}C_{\eta,\eta',j,j',k} JH^{j'}G_{\eta'}(JH^{j}G_{\eta})^\dag.
\end{equation}
Here, $\eta$ encompasses all $d$-dimensional spherical harmonic indices, $\ell$ denotes the first spherical harmonic index\footnote{Even in two dimensions, I say $\ell$ for consistency. Since $\ell$ can be negative in two dimensions, I must write $|\ell|$ here and some other points in this work.}, and $j$ is a nonnegative integer. $\dag$ denotes the standard Hermitian adjoint.  $G_{\eta}$ is a vector in $\mathbb{C}[x_1,\dots,x_d]^2$ defined as,
\begin{equation}\label{eq:Gdefinition}
G_\eta(\vectorbold x)=\begin{bmatrix}
    0\\r^\ell Y_\eta(\bm{\theta})
\end{bmatrix}.
\end{equation}
The function $r^\ell Y_\eta(\bm{\theta})$ is in fact a polynomial. To compute the Cartesian form of these polynomials, see Appendix \ref{sec:harmPolys}. 
Also appearing in Equation \ref{sigmaDecomposition} is $H$, the linear map on $\mathbb{C}[x_1,\dots,x_d]^2$ given by $\frac{1}{\pi}\arccot(J\sigma_0)$ or alternatively,
\begin{equation}\label{eq:hDefinition}
H\begin{bmatrix}\phi_1\\\phi_2\end{bmatrix}=\frac{1}{2}\begin{bmatrix}
    \nabla\cdot(p\nabla\phi_2)-(d-1)\phi_2\\p\phi_1
\end{bmatrix}.
\end{equation}
The equivalence of these two definitions comes from the expression of the modular Hamiltonian of the vacuum on the ball. \cite{Casini_2010}
Since $\alpha^{(k)}$ is finite rank, we will see that computing this decomposition amounts to computing a finite dimensional matrix inverse. By inserting this decomposition, the trace $\Trace \left[\prod_{q=1}^n (z-J\sigma_0)^{-1}J\sigma^{(k_q)}\right]$ becomes a finite linear combination of products of terms of the form,
\begin{equation}
   (JH^{j_2}G_{\eta_2})^\dag (z-J\sigma_0)^{-1} H^{j_1}G_{\eta_1}  
\end{equation} 
But, since $(z-J\sigma_0)^{-1} $ commutes with rotations of the ball, Schur's Lemma \cite{Hall:371445} implies that this is only nonzero if $\eta_1=\eta_2$ and that it only depends on $\ell$, the first index of $\eta$. I therefore define the \textit{matrix element} in $d$ spatial dimensions as,
\begin{equation}\label{eq:matrixElemDfn}
    M_{d,\ell,j_1,j_2}(z)=(JH^{j_2}G_{\eta_\ell})^\dag (z-J\sigma_0)^{-1} H^{j_1}G_{\eta_\ell},
\end{equation} 
where $\eta_\ell$ is the spherical harmonic index $(\ell,0,\dots,0)$.
So, to compute the trace in Equation \ref{eq:GeneralDecompisition} one needs to compute the relevant matrix elements, which I do in Section \ref{sec:matrixElems} and the decomposition in Equation \ref{sigmaDecomposition}, which I describe how to do in \ref{sec:FiniteTrace}.

\subsection{The Form of the Matrix Elements}\label{sec:matrixElems}
We begin by computing the matrix elements defined in Equation \ref{eq:matrixElemDfn}. First, we make an important observation. Note that $H^\dag=JHJ$ and that, since $H=\frac{1}{\pi}\arccot(J\sigma_0)$, $H$ commutes with the resolvent $(z-J\sigma_0)^{-1}$. This allows one to derive the identity,
\begin{equation}
    M_{d,\ell,j_1,j_2}(z)=(-1)^{j_2}M_{d,\ell,j_1+j_2,0}(z).
\end{equation}
 In other words, up to a sign, the matrix elements only depends on $j\coloneq j_1+j_2$. It suffices now to compute \begin{equation}
     M_{d,\ell,j,0}(z) = (JG_{\eta_\ell})^\dag (z-J\sigma_0)^{-1} H^{j}G_{\eta_\ell}.
 \end{equation}
In \cite{firstPaper}, we derived the eigendecomposition of the differential operator $H = \frac{1}{\pi}\arccot(J\sigma_0)$ with the transformation into the eigenbasis involving a continuous Fourier-like transform. In this eigenbasis, we can write the matrix element as a single integral:
\begin{equation}
     M_{d,\ell,j,0}(z)=\frac{i}{\Gamma(\gamma)^2} \int_{-\infty}^{\infty}\diff u \text{ } \frac{(iu)^j\csch(\pi u)|\Gamma(\gamma+iu)|^2}{z+i\coth(\pi u)}.
\end{equation}
Here, $\gamma = \frac{d-1}{2}+|\ell|$. This integral is computed in Appendix \ref{sec:matrixElemsApp}. If $\gamma$ is an integer, the result is, \begin{equation}
    M_{d,\ell,j,0}(z) =  \frac{\pi}{\gamma\Gamma(\gamma)^2}P_{\gamma,j}(\frac{\arctan(z)}{\pi}), 
\end{equation}where $P_{\gamma,j}$ is a rational polynomial satisfying $P_{\gamma,j}(\frac{1}{2})=P_{\gamma,j}(-\frac{1}{2})=0$. If $\gamma$ is a half-integer, the result is, \begin{equation}
    M_{d,\ell,j,0}(z) =  \frac{\pi}{\gamma\Gamma(\gamma)^2z}P_{\gamma,j}(\frac{\arctan(z)}{\pi}), 
\end{equation} where $P_{\gamma,j}$ is a rational polynomial satisfying $P_{\gamma,j}(0)=0$. We can write $P_{\gamma,j}(z)$ as
\begin{equation}
    P_{\gamma,j}(x) = P_{\gamma,0}(x)\sum_{k=0}^j\frac{2\gamma}{2\gamma+k}\left[\sum_{m=0}^{j-k}(-1)^{j-k} {j \choose m} \stirling{j-m}{k}\gamma^m\right]\left(\gamma+\frac{1}{2}+x\right)^{(k)}
    \end{equation}
where, $t^{(k)}=\prod_{n=0}^{k-1}
(t+n)$. If $\gamma$ is an integer, $ P_{\gamma,0}$ is,
\begin{equation}
    P_{\gamma,0}(x) =  \prod_{k=1}^\gamma ((k-\frac{1}{2})^2-x^2),
\end{equation}
and if $\gamma$ is a half integer $P_{\gamma,0}$ is,
\begin{equation}
    P_{\gamma,0}(x) = x\prod_{k=1}^{\gamma-\frac{1}{2}} (k^2-x^2).
\end{equation}

\subsection{Practically Computing the Traces}\label{sec:FiniteTrace}
In this section, I describe practical methods for computing the trace in Equation \ref{GeneralDecompisition}, with the aim to present a conceptually simpler strategy than that of \cite{firstPaper}, which computes the trace with a sum of sequences of spherical harmonic indices. Instead, we represent this trace as a trace of a finite dimensional matrix. For this section, I will only deal with the most general case. Symmetry can reduce the dimensionality of the matrices involved significantly, which will be discussed in Section \ref{symmetry}.

For each $\sigma^{(k)}$, we construct a square matrix $A_k$ with dimension equal to the size of this set whose entries are  the coefficients in Equation \ref{sigmaDecomposition}:
\begin{equation}
    (A_k)_{p,q} =C_{\eta_p,\eta_q,j_p,j_q,k}, 
\end{equation}
where $p$ and $q$ index all pairs $(\eta,j)$ of spherical harmonic indices $\eta$ (with the first index denoted by $\ell$) and nonnegative integers $j$ such that $|\ell|+j\leq  \max_q(k_q)$. 
Here $p$ and $q$ indexes the rows and columns of the matrix, respectively. Practically, we can compute the $A_k$ by first constructing the matrices $\bar{A}_k$ representing $\sigma^{(k)}$ in the standard monomial basis and a change of basis matrix $P$ whose columns are the components of the basis defined in Section \ref{polybasis} written in the standard monomial basis. From Equation \ref{sigmaDecomposition}, we can write, $\bar{A}_k = PA_kP^\dag$, which can be inverted to get,
\begin{equation}
    A_k = P^{-1}\bar{A}_k(P^{-1})^{\dag}.
\end{equation}
We also construct a matrix $B$ composed of the matrix elements computed in Section \ref{sec:matrixElems},
\begin{equation}
    B_{p,q} = \delta _{\eta_p,\eta_q}M_{d,\ell_p,j_p,j_q}(z).
\end{equation}
Note that many of the entries in the $A_k$ and $B$ matrices will be zero.
We can now rewrite the trace appearing in Equation \ref{GeneralDecompisition} as a trace involving the finite dimensional matrices $A_k$ and $B$,
    \begin{equation}\label{GeneralDecompisition}
         S_\alpha^{[N]} = \sum_{\substack{N=\sum_{j=1}^nk_j \\n\geq 1,\text{ } k_j\geq 1}}\frac{(-1)^n}{2\pi n i }\oint_\xi \diff z \text{ }h_\alpha(z)\Trace \left[\prod_{q=1}^n BA_{k_q}\right].
\end{equation}
The $(-1)^n$ comes from a factor of $J^2$ appearing in front of each matrix element.
Alternatively, one can compute all contributing traces up to a fixed order $N_{max}$ at once by combining the $A_{k}$ into a single matrix $A(t)=\sum_{k=1}^{N_{max}}A_kt^k$\footnote{For mutual information, you can set the upper limit of the sum to $\frac{N_{max}}{2}$ because the $n=1$ term won't contribute.}. Equation \ref{GeneralDecompisition} becomes, 
\begin{equation}\label{GeneralDecompositionFixed Order}
         \sum_{N=1}^{N_{max}}S_\alpha^{[N]}t^N + \mathcal{O}(t^{N_{max}+1})  = \sum_{\substack{N=\sum_{j=1}^nk_j \\n\geq 1,\text{ } k_j\geq 1}}\frac{(-1)^n}{2\pi n i }\oint_\xi \diff z \text{ }h_\alpha(z)\Trace \left[( BA(t))^n\right].
\end{equation}
Note the similarity to Equation \ref{eq:GeneralDecomposition_n}.
So, we simply have to compute the right hand side and appropriately truncate. This approach is simpler to code at the expense of some unnecessary computation of incomplete higher order terms. Regardless, the extra computation does not add much time and so this latter approach was used to compute the results in this paper.

\subsection{Taking Advantage of Symmetry}\label{symmetry}
Here, I discuss how to take advantage of a spherically symmetric system. Of our examples, a ball at finite temperatures is spherically symmetric but two balls separated by a distance $R$ only obeys cylindrical symmetry around the axis connecting the ball's centres. I focus here only on spherical symmetry here, as the case of cylindrical symmetry is more complicated and only significantly helps for high dimensions. For more details, including on using cylindrical symmetry, see \cite{firstPaper}.
Spherical symmetry causes each $\sigma^{k}$ to commute, as a linear map of 2-vector fields on the ball, with rotation of the input. Schur's Lemma \cite{Hall:371445} therefore implies the coefficients in Equation \ref{sigmaDecomposition} are zero if $\eta \neq \eta'$ and that the coefficients only depend on the outermost spherical harmonics index $\ell=\ell'$ along with $j$, $j'$ and $k$. 
Recall that $\eta_\ell$ is the spherical harmonic index $(\ell,0,\dots,0)$. Let $P_\ell$ be the projector onto the subspace of functions of the form $f(r)Y_{\eta_\ell}(\bm{\theta})$:

\begin{equation}
    P_{\ell}[\phi](r,\bm{\theta}) =Y_{\eta_\ell}(\bm{\theta})\int_{\partial\Omega}\diff^{d-1}\bm{\alpha}\text{ } \phi(r,\bm{\alpha})Y_{\eta_\ell}(\bm{\alpha}).
\end{equation}
Here, $r$ is a radial coordinate, both $\bm{\alpha}$ and $\bm{\theta}$ are vectors encompassing all angular variables, As well, $\partial\Omega$ denotes the surface of the unit ball. We multiply Equation \ref{sigmaDecomposition} on the left by $P_\ell$. The result reduces the size of the sum significantly,
\begin{equation}\label{eq:reduceddecomp}
   P_\ell \sigma^{(k)}= \sum_{\substack{j,j'\\j+j' =k-d-2\ell+1}}C_{\eta_{\ell},\eta_{\ell},j,j',k} JH^{j'}G_{\eta_{\ell}}(JH^{j}G_{\eta_{\ell}})^\dag.
\end{equation}
If we can explicitly compute the left hand side, then we have reduced the dimensionality of computing the decomposition in Equation \ref{sigmaDecomposition} by a large amount. Specifically, we need the kernel of $P_\ell \sigma^{(k)}$, whose entries must be writable in the form,
\begin{equation}
K(\vectorbold{x},\vectorbold{y}) = W^{q}_\ell(r,s)Y_{\eta_\ell}(\bm{\alpha}) Y_{\eta_\ell}(\bm{\theta}).     
\end{equation}
Here $\vectorbold{x}=(r,\bm{\theta})$ and $\vectorbold{y}=(s,\bm{\alpha})$ in spherical coordinates and $W^{q}_\ell(r,s)$ must be a homogeneous two-variable polynomial. 

For the case of thermal entropy, the entries of the kernel of $\sigma^{k}$ must have the form $C|\vectorbold{x}-\vectorbold{y}|^{2q}$ for some constant $C$ and nonnegative integer $q$ (c.f. Equation \ref{eq:entKernel}). This is a consequence of the \textit{translational} as well as rotational invariance of the wider field theory on all space. Proceeding further in this case, to compute $P_\ell\sigma^{(k)}$, it suffices to compute the integral, 
\begin{equation}
    W^{q}_\ell(r,s)Y_{\eta_\ell}(\bm{\alpha}) = \int_{\partial \Omega} \diff^{d-1}\bm{\theta}\text{ } |\vectorbold{x}-\vectorbold{y}|^{2q} Y_{\eta_\ell}(\bm{\theta}).
\end{equation}
I do so in appendix \ref{sec:radialKernel}. The result is, 
\begin{equation}
   W^{q}_\ell(r,s)=2q!\pi^{\frac{d}{2}} \sum_{\rho=0}^{\lfloor\frac{q-\ell}{2}\rfloor} \frac{(-rs)^{2\rho+\ell}(r^2+s^2)^{q-2\rho-\ell}}{\rho!(q-\ell-2\rho)!\Gamma(\frac{d}{2}+\rho+\ell)}.
\end{equation}
After computing $P_\ell \sigma^{(k)}$, we can solve a $k-2l$ dimension linear system to find the $C_{\eta_{\ell},\eta_{\ell},j,j',k}$ thanks to Equation \ref{eq:reduceddecomp}. This significantly reduces the computation time compared to solving the constants from the full expansion \ref{sigmaDecomposition}.

One other way spherical symmetry improves the speed of this computation is that, due to Schur's Lemma \cite{Hall:371445}, each $A_k$ defined in Section \ref{sec:FiniteTrace} can be decomposed as a direct sum of block matrices. Specifically, let $A_{k,\ell}$ be the matrix with entries $(A_{k,\ell})_{j,j'}=C_{\eta_\ell,\eta_\ell,j,j',k}$. Then,
\begin{equation}
    A_{k} = \bigoplus_{|\ell|\leq \frac{k-d+1}{2}} A_{k,\ell}^{\oplus \left[N(d,\ell)\right]},
\end{equation}
where $A^{\oplus n}$ denotes $A$ direct summed with itself $n$ times and $N(d,\ell)=\binom{d + l - 1}{d - 1} - \binom{d + l - 3}{d - 1}$ is the number of spherical harmonics in $d$ spatial dimensions with first index $\ell$. \cite{frye2012spherical}
An analogous decomposition exists for $B$ into the block matrices $B_\ell$ with entries $(B_{\ell})_{j,j'}=M_{d,\ell_\ell,j,j'}(z)$,
This decomposition allows us to write the trace in Equation \ref{GeneralDecompisition} as a sum of traces of block matrices,
    \begin{equation}\label{eq:SymmetrizedTrace}
\Trace \left[\prod_{q=1}^n BA_{k_q}\right]=\left[N*d,\ell)\right]\sum_\eta\Trace \left[\prod_{q=1}^n B_\eta A_{k_q,\eta}\right].
\end{equation}
Both these results allow for significant improvements to the speed of the computation which allows computation of the first five nonzero terms of the thermal entropy difference in up to six spatial dimensions.

\subsection{Computing the Contour Integral}\label{sec:contourInts}
Now that we can compute the traces appearing in Equation \ref{eq:GeneralDecompisition}, we need to compute the resulting contour integral over the contour shown in Figure \ref{fig:double_keyhole}. Throughout this section, I use the Bernoulli numbers and polynomials described in Appendix \ref{sec:bernoulli}.

In odd dimensions, the contour integral for each $S_\alpha^{[N]}$ has the form,
\begin{equation}\label{eq:contourFormOdd}
     S_\alpha^{[N]}=\frac{1}{2\pi}\int_{\xi} \diff z\text{ } \frac{\alpha}{2(\alpha-1)}W(\arctan(z))\left[\frac{e^{2(\alpha-1) i \arccot(z)}-1}{(1-iz)e^{2(\alpha-1) i \arccot(z)}+(1+iz)}\right],
\end{equation}
where $W(x)=\sum_{j=0}^{\deg(W)}w_jx^j$ is some polynomial satisfying $W(\frac{\pi}{2})=0$. In Appendix \ref{sec:oddDimCalc}, I compute this contour integral as,
\begin{equation}
    S_\alpha^{[N]}= \sum_{j=0}^{\deg(W)}w_j\left(\frac{\pi}{\alpha}\right)^j 
    \left[\frac{B_{j+1}(\frac{\alpha}{2})}{(j+1)(1-\alpha)}\right].
\end{equation}
In particular, for Von Neumann entropy, we have,
\begin{equation}
   S_1^{[N]}= \frac{1}{2}\sum_{j=0}^{\deg(W)}w_j\pi^j 
    (1-2^{1-j})B_j.
\end{equation}
In even dimensions, the integral is a linear combination of terms of the form
\begin{equation}\label{eq:contourFormEven}
    I_{j,n,\alpha}=\frac{1}{2\pi}\int_{\xi} \diff z\text{ } \frac{\alpha}{2(\alpha-1)}\frac{\arctan(z)^j}{z^n}\left[\frac{e^{2(\alpha-1) i \arccot(z)}-1}{(1-iz)e^{2(\alpha-1) i \arccot(z)}+(1+iz)}\right],
\end{equation}
where $j\geq n\geq 1$. This integral is more difficult to compute. In the remainder of Appendix \ref{sec:RenyiInts}, I compute it only for $\alpha \in \mathbb{Z}_{\geq 0}$ and $\alpha \to \infty$. For $\alpha = 0$, this integral is divergent. 
For $\alpha = 1$ and $n=1$, it is
\begin{equation}
    I_{j,1,1}= \frac{1}{j+1}(2^{-j-1}-1)B_{j+1},
\end{equation}
For $\alpha=1$ and  $n > 1$, it is instead,
\begin{equation}\label{eq:jnEvena=1}
     I_{j,n,1}=-\frac{1}{2(j+1)}\left[\sum_{k=\frac{j-n}{2}+1}^{\lfloor\frac{j}{2}\rfloor}\pi^{2k}{j+1\choose 2k}B_{2k} \sum_{\substack{\sum_{l=1}^{j+1-2k} q_l = n-1\\q_l \text{ odd}}}\left(\prod_l\frac{1}{q_l}\right)\right].
\end{equation}
If $\alpha$ is an integer greater than one, the contour integral is
\begin{equation}\label{eq:aIntegerfinal}
    I_{j,n,\alpha}=\frac{1}{2(1-\alpha)} \sum_{j=1}^{\alpha-1} \tan(\frac{j\pi}{\alpha})^n (\frac{\pi}{2}-\frac{j\pi}{\alpha})^n.
\end{equation}
Lastly, for $\alpha = \infty$, there is no simple explicit expression for the integral. Instead, I present a recurrence relation,
\begin{equation}
    I_{j,n,\infty} = (\frac{m}{n-1})I_{j-1,n-1,\infty}-I_{j,n-2,\infty}.
\end{equation} The base cases are
\begin{equation}
    I(2m,0)=-\frac{1}{2(2m+1)}\left(\frac{\pi}{2}\right)^{2m},
\end{equation}
and,
\begin{equation}
    I(2m+1,1) = \frac{(2j+1)!}{2^{2j+1}}\sum_{k=0}^j \frac{(-1)^k\pi^{2(j-k)}(4^{-k}-1)}{(2j-2k+1)!}\zeta(2k+1).
\end{equation}
Here $\zeta(s)=\sum_{j=1}^\infty\frac{1}{j^s}$ is the Riemann zeta function. 
Together, with our results about the matrix elements these integrals allow us to compute the $S_\alpha^{[N]}$ for all $\alpha$ in odd dimensions and for $\alpha \in \mathbb{Z}_{\geq 0}$ and $\alpha =\infty$ in even dimensions. I note that a consequence of this work is that $S_\alpha^{[N]}$ always \textit{exists} for all $N$ if our conditions outlined at the beginning of this section are satisfied, showing that $ S_\alpha(\hat\rho(t))-S_\alpha(\hat\rho(0))$ is almost certainly analytic.\footnote{In even spatial dimensions and non-integer $\alpha>0$, the contour integral in Appendix \ref{sec:renyi} always converges, showing that the coefficients are always finite, even without an analytic result.} \footnote{Strictly speaking, since the $S_\alpha^{[N]}$ were computing by taking derivatives, I have shown it is infinitely differentiable at zero, but not necessarily analytic. If it were not analytic, the series presented in this paper would not converge anywhere except at $t=0$. This happens if the terms of the series grow too fast. While this seems unlikely, especially with numerical results like \cite{Chen_2017}, the possibility is worth keeping in mind. } .

\section{The Expansions for Mutual Information of Distant Balls}\label{sec:MI}
Consider two balls of radius $r$ in a massless scalar vacuum separated by a distance $R$. The goal of this section is to compute the first few terms of the expansion for the Rényi mutual information between the balls for dimensions two through six,
\begin{equation}
   I_{d,a}(r,R)= \sum_{N=1}^\infty I_{d,\alpha}^{[N]}\left(\frac{r}{R}\right)^{N}.
\end{equation} 
The simplest approach to this calculation uses a result from \cite{firstPaper} which takes advantage of the identicalness of the two balls. In this case, the expression for $I_{d,\alpha}^{[N]}$ is identical to Equation \ref{eq:GeneralDecomposition_n} except there is a factor of $-2$ and only the terms with even powers contribute 
\begin{equation}
      I_{d,\alpha}(r, R) = -\sum_{n=1}^\infty\frac{1}{\pi n i }\oint_\xi \diff z \text{ }h_\alpha'(z)\Trace \left[\left((z-J\sigma_0)^{-1}J\delta\sigma\right)^{2n} \right].
\end{equation}
Equivalently, we can write the individual coefficients of the expansions as 
\begin{equation}\label{eq:GeneralDecompisitionMI}
         I_{d,\alpha}^{[N]} = -\sum_{\substack{N=\sum_{j=1}^{2n}k_j \\n\geq 1,\text{ } k_j\geq 1}}\frac{1}{\pi n i }\oint_\xi \diff z \text{ }h_\alpha(z)\Trace \left[\prod_{q=1}^{2n} (z-J\sigma_0)^{-1}J\sigma^{(k_q)} \right].
\end{equation}
This allows us to use the above results with only slight modification.
Following the convention of the above sections the $\sigma^{(k)}$ are the terms of a Taylor series,
\begin{equation}
   \delta \sigma =  \sum_{k=1}^\infty \sigma^{(k)} R^{-k}
\end{equation}
where $\delta \sigma$ is a linear map with an integral kernel 
\begin{equation}\label{eq:miKernel}
   K(\vectorbold{x},\vectorbold{y};\frac{1}{R}) = \frac{\Gamma(\frac{d-1}{2})}{2\pi^{\frac{d+1}{2}}} \begin{bmatrix}
      \frac{1}{|\vectorbold{R}-(\vectorbold{y}+\vectorbold{x})|^{d-1}}&0\\ 0&-\frac{d-1}{|\vectorbold{R}-(\vectorbold{y}+\vectorbold{x})|^{d+1}}
    \end{bmatrix},
\end{equation}
where $\vectorbold{R}=(R,0,\dots,0)$. This follows from the forms of the two-point correlation functions in Equation \ref{eq:themalCorrelators} and formulae for powers of the Laplacian well known in literature of pseudo-differential equations \cite{Stinga2018UsersGT}. As we can see, this the entries of this kernel are analytic in $\frac{1}{R}$, $\vectorbold{x}$, and $\vectorbold{y}$ and the expansion parameter $\frac{1}{R}$ has units of energy. Therefore, the assumptions needed for the results of Section \ref{sec:entropyExpans} to apply are valid. To compute the polynomials kernels of the first few $\sigma^{(k)}$, note that one only needs to compute the corresponding terms of the series of $\frac{1}{((z-R)^2+s^2)^p}$, which is straightforward in any computer algebra system.  

To compute the terms of Rényi mutual information, I use the procedure in Section \ref{sec:FiniteTrace} but modify the right hand side of Equation \ref{GeneralDecompositionFixed Order} to only include even powers and include a factor of $-2$. I present here the final results for dimensions two through six up to three nonzero terms. I highlight the results for Von Neumann entropy.
In two dimensions the result is,
\begin{align}\label{eq:MI2}
    I_{2,1}(r, R) &= \frac{1}{3} \left(\frac{r}{R}\right)^2 + \left(\frac{26}{45}+\frac{4}{3 \pi^{2}}\right) \left(\frac{r}{R}\right)^4 +  \left(\frac{608}{315}+\frac{40}{9 \pi^{2}}+\frac{16}{3 \pi^{4}}\right) \left(\frac{r}{R}\right)^{6} + \mathcal{O}\left(\left(\frac{r}{R}\right)^{8}\right) \\
    &= 0.333 \left(\frac{r}{R}\right)^2 + 0.713 \left(\frac{r}{R}\right)^4 + 2.435 \left(\frac{r}{R}\right)^{6} + \mathcal{O}\left(\left(\frac{r}{R}\right)^{8}\right).
\end{align}
In three dimensions, the expansion is,
\begin{align}
    I_{3,1}(r, R) &= \frac{4}{15} \left(\frac{r}{R}\right)^4 + \frac{32}{35} \left(\frac{r}{R}\right)^6 + \frac{416}{105} \left(\frac{r}{R}\right)^8 + \mathcal{O}\left(\left(\frac{r}{R}\right)^{10}\right) \\
                   &= 0.267 \left(\frac{r}{R}\right)^4 + 0.914 \left(\frac{r}{R}\right)^6 + 3.962 \left(\frac{r}{R}\right)^8 + \mathcal{O}\left(\left(\frac{r}{R}\right)^{10}\right).
\end{align}
In four dimensions, the result is,
\begin{align}
    I_{4,1}(r, R) &= \frac{8}{35} \left(\frac{r}{R}\right)^6 + \frac{128}{105} \left(\frac{r}{R}\right)^8 + \frac{4096}{693} \left(\frac{r}{R}\right)^{10} + \mathcal{O}\left(\left(\frac{r}{R}\right)^{12}\right) \\
    &= 0.229 \left(\frac{r}{R}\right)^6 + 1.219 \left(\frac{r}{R}\right)^8 + 5.911 \left(\frac{r}{R}\right)^{10} + \mathcal{O}\left(\left(\frac{r}{R}\right)^{12}\right).
\end{align}
In five dimensions, it is,
\begin{align}
    I_{5,1}(r,R)&=\frac{64}{315} \left(\frac{r}{R}\right)^{8}+ \frac{1024}{693} \left(\frac{r}{R}\right)^{10}+\frac{25600}{3003} \left(\frac{r}{R}\right)^{12}+ \mathcal{O}\left(\left(\frac{r}{R}\right)^{14}\right)\\
               &=0.203 \left(\frac{r}{R}\right)^{8}+ 1.478 \left(\frac{r}{R}\right)^{10}+8.525 \left(\frac{r}{R}\right)^{12}+ \mathcal{O}\left(\left(\frac{r}{R}\right)^{14}\right).
\end{align}
Lastly in six dimensions, it is,
\begin{align}
    I_{6,1}(r,R)&=\frac{128}{693} \left(\frac{r}{R}\right)^{10}+ \frac{5120}{3003} \left(\frac{r}{R}\right)^{12}+\frac{8192}{715} \left(\frac{r}{R}\right)^{14}+ \mathcal{O}\left(\left(\frac{r}{R}\right)^{16}\right)\\
               &=0.185 \left(\frac{r}{R}\right)^{10}+ 1.705 \left(\frac{r}{R}\right)^{12}+11.457 \left(\frac{r}{R}\right)^{14}+ \mathcal{O}\left(\left(\frac{r}{R}\right)^{16}\right).
\end{align}
The lowest order terms in each dimension each agree with \cite{Cardy_2013,Agon2015,Chen_2018,firstPaper}. The next to leading order term in two dimensions agrees with \cite{Chen_2017}. In addition, all terms presented in three and five dimensions agree with \cite{chen_odd}. I also note that, while it appears that only $d=2$ has irrational coefficients, Equation \ref{eq:jnEvena=1} tells us that in any even dimension inverse powers of $\pi$ start appearing when terms with $n=4$ contribute to Equation \ref{eq:GeneralDecompisitionMI}. This happens at the $4(d-1)$-th power and beyond.

See Appendix \ref{sec:RenyMIFull} for expansions for more general Rényi parameters $\alpha$, where I give the first few terms of the expansion for all $\alpha$ in odd dimensions and for $\alpha\in\{2,3,4,\infty\}$ in even dimensions.

\section{The Expansions for Thermal Entropy at Low Temperature}\label{sec:Ent}
Consider a ball of radius $r$ in a thermal massless scalar field at temperature $T$ in $d$ spatial dimensions. In this section, I describe how to compute the following expansion for the entropy difference. 
\begin{equation}
  S_{d,\alpha}(r,T)- S_{d,\alpha}(r,0) = \Delta S_{d,\alpha}(r,T)= \sum_{N=1}^\infty S_{d,\alpha}^{[N]}\left(\frac{r}{R}\right)^{N},
\end{equation} 
for dimensions three through six up to five nonzero terms.
We again use the convention that the $\sigma^{(k)}$ are the terms of a Taylor series,
\begin{equation}
   \delta \sigma =  \sum_{k=1}^\infty \sigma^{(k)} T^{k}
\end{equation}
where the $\sigma^{(k)}$ are linear maps whose kernels were derived in \cite{firstPaper}. These kernels are only nonzero if $k-d+1$ is even, in which case they are,
\begin{equation}\label{eq:entKernel}
   K^{(d-1+2j)}(\vectorbold{x},\vectorbold{y}) = \frac{(-1)^j(d-2+2j)!\zeta(d-1+2j)(d-2){!}{!}}{2^{d-2}\pi^{\frac{d}{2}}\Gamma(\frac{d}{2}) (2j){!}{!}(d+2j-2){!}{!}}\begin{bmatrix}
    |\bm{x}-\bm{y}|^{2j}&0\\ 0&  2j(2j+d-2)|\bm{x}-\bm{y}|^{2j-2}
    \end{bmatrix},
\end{equation}
where $\zeta(z)$ is the Riemann zeta function. We once again see that the entries of $\delta \sigma$ are analytic and that $T$ has units of energy, validating our assumptions outlined in Section \ref{sec:entropyExpans}. Computing the $S_{d,\alpha}^{[N]}$ is a direct application of Section \ref{sec:FiniteTrace}. However, for higher dimensions, the speed improvements described in Section \ref{symmetry} were necessary. Here I present the results for Von Neumann entropy for dimensions three through six. 

For $d=3$, we get:
\begin{align}
    \Delta S_{3,1}(r, T) &= \frac{1}{9} \pi^{2} (rT)^{2} + \frac{2}{675} \pi^{4}(rT)^4 + \frac{32}{8505} \pi^{6}(rT)^6 -\frac{272}{212625} \pi^{8}(rT)^8+ \frac{512}{1403325} \pi^{10}(rT)^{10}+ \mathcal{O}\left((rT)^{12}\right) \\
                   &= 1.097 (rT)^{4} + 0.289 (rT)^{4} + 3.617 (rT)^{6} -12.138 (rT)^{8}+36.167(rT)^{10}+\mathcal{O}\left((rT)^{12}\right).
\end{align}
The result for four spatial dimensions is,
\begin{align}
    \begin{split}
        \Delta S_{4,1}(r,T) &= \frac{3 \pi  \zeta \! \left(3\right)}{16}(rT)^3 + \frac{ \pi  \zeta \! \left(5\right)}{2}(rT)^{5}-\frac{16 \zeta \! \left(3\right)^{2} }{35}(rT)^6  + \frac{128 \zeta \! \left(3\right)^{3} (rT)^{9}}{45 \pi}-\frac{8192  \zeta \! \left(5\right)^{2}}{693}(rT)^{10}+\mathcal{O}\left((rT)^{11}\right)\\
        &=0.708(rT)^{3}+1.629(rT)^{5}+0.661(rT)^{6}+1.573(rT)^{9}-12.701(rT)^{10}+\mathcal{O}\left((rT)^{11}\right)
    \end{split}
\end{align}
For five dimensions it is,
\begin{align}
    \begin{split}
        \Delta S_{5,1}(r,T) &= \frac{4\pi^{4}}{675}  (rT)^{4} +\frac{32\pi^{6}}{19845} (rT)^{6} - \frac{32\pi^{8}}{637875} (rT)^{8} 
         - \frac{131072\pi^{12}}{7449316875} (rT)^{12} +\frac{65536\pi^{14}}{17239847625} (rT)^{14} \\&\quad + \mathcal{O}\left((rT)^{16}\right)\\
        &= 0.577 (rT)^{4} + 1.550 (rT)^{6} - 0.476 (rT)^{8} - 16.263 (rT)^{12} + 34.677 (rT)^{14}+ \mathcal{O}\left((rT)^{16}\right)
    \end{split}
\end{align}
Lastly, for six spatial dimensions it is.
\begin{align}
    \begin{split}
        \Delta S_{6,1}(r,T) &= \left(\frac{5 \pi \zeta \! \left(5\right)}{32}\right) (rT)^{5} + \left(\frac{15 \pi \zeta \! \left(7\right)}{32}\right) (rT)^{7} - \left(\frac{256 \zeta \! \left(5\right)^{2}}{693}\right) (rT)^{10} \\
        &\quad - \left(\frac{16384 \zeta \! \left(7\right)^{2}}{715}\right) (rT)^{14} + \left(\frac{4096 \zeta \! \left(5\right)^{3}}{945 \pi}\right) (rT)^{15} + \mathcal{O}\left((rT)^{16}\right)\\
        &= 0.509 (rT)^{5} + 1.485 (rT)^{7} - 0.397 (rT)^{10} - 23.299 (rT)^{14}+ 1.538 (rT)^{15}+ \mathcal{O}\left((rT)^{16}\right)
    \end{split}
\end{align}
The leading order terms are in agreement with \cite{firstPaper}. Interestingly, only the powers which are whole number combinations of $d-1$ and $d+1$ appear. This is not true for other values of $\alpha$ whose series are given in Appendix \ref{sec:RenyEntull}. 

\section{Conclusion}\label{sec:conlusion}
Extending on the work of \cite{firstPaper}, I have described a strategy to compute the perturbation series for the Rényi entropy of a quantum state on a ball assuming the perturbed state is the vacuum state of a massless scalar, the perturbation parameter has units of energy and the perturbations in the two point correlation functions are analytic. Under these conditions, I showed that the coefficients are finite for all orders and outline a procedure to compute the leading order terms of the series for any fixed dimensions. I then used this procedure to compute the lowest order terms of the expansion for mutual information of distant balls in spatial dimensions two, three, four, five, and six. I also used this to compute the perturbation series for a thermal field at low temperature in spatial dimensions three, four, five, and six. I also computed the corresponding expansions for Rényi entropy for all Rényi parameters $\alpha$ in odd spatial dimensions and for integer $\alpha$ and $\alpha=\infty$ in even spatial dimensions. 

It is worth pointing out to the interested reader that the series expansion can also be computed analytically for any rational $\alpha$ in even dimensions, not only integers, by solving the corresponding contour integral in Appendix \ref{sec:contourInts}. I neglect it in this work solving it due to the complexity of the result and the unlikeliness that the result is of any interest. 

One potential refinement of this work would be to discuss how cylindrical symmetry could reduce the time needed to calculate the expansion for mutual information. Any interested reader reproducing these calculations will observe that the computation time becomes increasingly intractable for dimensions larger than six. Generalizing the discussion of symmetry in this work from spherical to cylindrical symmetry would reduce this computation time.

This work naturally extends to expansions where there are multiple expansion parameters with units of energy (or positive integer powers of energy). Interesting examples of this would be the entropy difference of a thermal field, expanding in $T$ and $T$, the mutual information of distant balls in a massive scalar field at vacuum, expanding in $m$ and $R^{-1}$; the mutual information of in a massless thermal field, expanding in $T$ and $R^{-1}$; or the mutual information of a massive thermal field, expanding in $T$, $m$ and $R^{-1}$. 

Much of the the work in this paper is applicable to numerical work as well. For example, suppose one wanted to numerically compute the coefficients of the expansion of standard mutual information of distant regions of arbitrary shapes, say cubes. The simplest approach would likely be directly substituting a lattice discretization of the correlation matrix into Equation \ref{eq:GeneralDecompisitionLattice} or the equivalent contour integral from $z=-i$ to $z=i$ and residue at $z=\infty$ described in Appendix \ref{sec:a=1Even}. Alternatively, one could follow the approach of this work more closely and reduce the computation of the traces in Equation \ref{eq:GeneralDecompisition} to a sum of matrix elements similar to Equation \ref{eq:matrixElemDfn}. Applying $(z-J\sigma)^{-1}$ amounts to solving a pseudo differential equation on the region of interest in $\mathbb{R}^d$. This is unlikely to have an analytical solution of the region is not a ball but it is in principle doable numerically. 

The results presented in this paper agree with preexisting results in the literature. But while previous literature used sophisticated methods such as replica tricks and $\frac{1}{n}$ expansions, this work's approach is directly generalized from calculations done in standard quantum mechanics. Approaching problems in holography and quantum information with the philosophy of this work may prove complimentary to more conventional methods.

\appendix

\section{Rényi Entropy of Gaussian States in Lattice Field Theory}\label{sec:renyi}
In this appendix, we derive an expression for the Rényi entropy in lattice field theory which generalizes naturally to the continuum limit. A Gaussian state is completely described by its correlation matrix. So, to approximate a field theoretic Gaussian state, a common approach is to approximate its correlation matrix (which is actually an infinite dimensional linear map) with a finite dimensional matrix. This is the approach used in \cite{Srednicki_1993, Shiba_2012,Chen_2017}. Therefore, to approximate the Rényi entropy in a full field theory, it suffices to compute the Rényi entropy for a Gaussian state with a finite dimensional correlation matrix.

To do so, we begin with the simplest case of a one dimensional harmonic oscillator with inverse temperature $\beta$ and frequency $\omega$. We want to find the Rényi entropy, defined in terms of the density operator $\rho$,
\begin{equation}
    S_\alpha(\hat\rho) = \frac{1}{1-\alpha}\ln(\Trace(\hat{\rho}^\alpha)),
\end{equation}
where the parameter $\alpha$ satisfies $0\leq \alpha \leq \infty$. If $\alpha \in \{0,1,\infty\}$, the Rényi entropy is defined by continuity.
Let $Z=\frac{1}{1-e^{-\beta\omega}}$ be the partition function when the the ground state is set zero energy.. We compute the Rényi entropy as,
\begin{align}
    \begin{split}
        S_\alpha&=\frac{1}{1-\alpha}\ln(\Trace\left(\left[\frac{e^{-\beta\omega\hat{N}}}{Z}\right]^\alpha\right))\\
        & =\frac{1}{1-\alpha}\ln\left(\frac{\sum_{n=0}^\infty e^{-\alpha\beta\omega n}}{Z^\alpha}\right)\\
        &= \frac{\ln(1-e^{-\alpha\beta\omega})}{\alpha-1}-\frac{\alpha\ln(1-e^{-\beta \omega})}{\alpha-1}
    \end{split}
\end{align}

\begin{figure}[htbp]
    \centering
    \begin{tikzpicture}[scale=1.2]
        \draw[ ->] (-3.2,0) -- (3.2,0) node[right] {$\Re$};
        \draw[ ->] (0,-3.2) -- (0,3.2) node[above] {$\Im$};
        
        \draw[thick,->] (-0.15, 1)-- (-0.15, 3);
        \draw[thick,<-] (0.15, 1)--(0.15, 3);
        \draw[thick,->] (-0.15, -1)-- (-0.15, -3);
        \draw[thick,<-] (0.15, -1)-- (0.15, -3);

         \draw[line width=0.8mm, dashed, blue] (0,-1) -- (0,1);
         \draw[line width=0.8mm, dashed, blue] (-3,-1) -- (3,-1);
        
          \node at (0,1)[circle,fill,inner sep=2pt]{};
          \node at (0.2,0.9) {\( i \)};  
        
          \node at (0,-1)[circle,fill,inner sep=2pt]{};
          \node at (-0.3,-0.8) {\( -i \)};  
          \node at (0,1.2)[circle,fill=red,inner sep=2pt]{};
          \node at (0,-1.2)[circle,fill=red,inner sep=2pt]{};
        
          \node at (0,2.7)[circle,fill=red,inner sep=2pt]{};
          \node at (0,-2.7)[circle,fill=red,inner sep=2pt]{};
        
          \node at (0,1.5)[circle,fill=red,inner sep=2pt]{};
          \node at (0,-1.5)[circle,fill=red,inner sep=2pt]{};
        
          \node at (0,2.3)[circle,fill=red,inner sep=2pt]{};
          \node at (0,-2.3)[circle,fill=red,inner sep=2pt]{};
        
          \node at (0,1.8)[circle,fill=red,inner sep=2pt]{};
          \node at (0,-1.8)[circle,fill=red,inner sep=2pt]{};

    \end{tikzpicture}
    \caption{A disconnected contour close to the portion of the imaginary line with magnitude greater than one. We are focused on the limiting case when the lines approach the portion of the imaginary axis with magnitude greater than one. Integration is taken in the direction of the arrows. Some example poles of $\Tr((z-J_D\sigma)^{-1})$ at the eigenvalues of $J_D\sigma$ are shown in red. The branch cuts and points of $h_\alpha$ depend on $\alpha$, but we show the branch for $\alpha = 1$ in blue as a representative example. For any $\alpha$, the branch cuts of $h_\alpha$ will never intersect a contour sufficiently close to the imaginary line.  }
    \label{fig:double_keyhole4}
\end{figure}
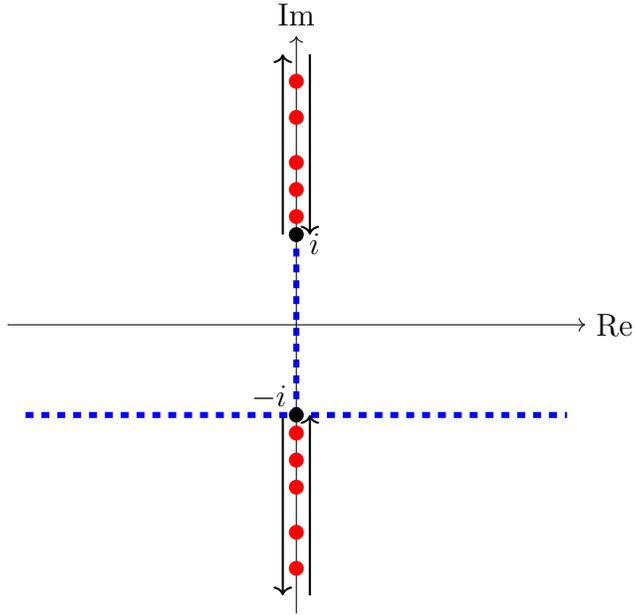

We now generalize this to an $D$ dimensional Gaussian state. To do so, it is convenient to define the matrix
\begin{equation}
    J_D=\begin{bmatrix} 0&I_D\\-I_D&0 \end{bmatrix}.
\end{equation} 
The Rényi entropy can be expressed in terms of the symplectic eigenvalues of the correlation matrix, which are the eigenvalues of what we call the skewed correlation matrix $J_D\sigma$. The symplectic eigenvalues have the form $\pm \lambda_k i$ where $\lambda_k\geq1$ and $k$ ranges from $1$ to $D$. The Gaussian state described by $J_D\sigma$ can be decomposed into a tensor product of $D$ thermal harmonic oscillators with $\beta_k=2\arccoth(\lambda_k)$ and $\omega_k=1$ \cite{martínmartínez2024quantummechanicsphasespace,firstPaper}. So by the additivity of Rényi entropy, this Gaussian state has Rényi entropy 
\begin{equation}
    S_\alpha=\sum_{k=1}^D \left[\frac{\ln(1-\left(\frac{\lambda_k-1}{\lambda_k+1}\right)^\alpha)}{\alpha-1}-\frac{\alpha\ln(1-\frac{\lambda_k-1}{\lambda_k+1})}{\alpha-1}\right].
\end{equation}
We express this entropy in terms of the trace of a matrix function, \footnote{It is not holomorphic in the neighborhood of $z=\pm i $. But, $\lambda=1$ corresponds to a zero temperature mode which does not contribute to the entropy since $h_\alpha(\pm i)=0$, letting us exclude such a mode from the calculation}
\begin{equation}
    S_\alpha = \Trace\left[h_\alpha(J_D\sigma)\right],
\end{equation}
where $h_\alpha$ is the holomorphic function,
\begin{equation}\label{eq:h_alphaApp}
   h_\alpha(z) = \frac{1}{4(\alpha - 1)} \Log \left( \left(1 - \left( \frac{z - i}{z + i} \right)^\alpha \right)^2 \right) + \frac{\alpha}{4(\alpha - 1)} \Log \left( -\frac{1}{4} (z + i)^2 \right).
\end{equation}
Note that $\Log(z)$ and $z^\alpha$ are taken to have branch cut on the negative real axis. Taking the limit $\alpha \to 1$, gives $h_1(z)=\frac{z-i}{2}\arccot(z)+\frac{1}{4} \Log \left( -\frac{1}{4} (z + i)^2 \right)$, which agrees with the function in Equation \ref{eq:matFun} for $|Im(z)|>1$. Since this function is holomorphic on a neighborhood of $\{z\in \mathbb{C}\mid  |\Im(z)+\frac{1}{2}| > \frac{1}{2}, \Re(z) = 0  \}$, which contains all possible eigenvalues of $J_D\sigma$, we can write the Rényi entropy as the following contour integral:
\begin{equation}\label{eq:contourEntropyApp}
     S_\alpha = \frac{1}{2\pi i }\oint_\xi \diff z \text{ }h_\alpha(z)\Tr((z-J_D\sigma)^{-1}).
\end{equation}
Here $\xi$ is the disconnected contour shown in Figure \ref{fig:double_keyhole4} in the limit of the contours approaching the imaginary axis. This is a consequence of the residue theorem and the fact that $h_\alpha(\pm i)=0$.
For the expansion used in this paper, we need the derivative of $h_\alpha$ which is the following odd function,
\begin{equation}\label{eq:gaderivativeApp}
h_\alpha'(z) = \frac{\alpha}{2(\alpha-1)}\left[\frac{e^{2i(\alpha-1)\arccot(z)}-1}{(z+i)e^{2i(\alpha-1)\arccot(z)}-(z-i)}\right],
\end{equation}
where $\arccot(z)$ is the principal branch.
In particular, the value of $h_\alpha'(z)$ for $|\Im(z)|>1$ is,
\begin{equation}
    h_\alpha'(it) = -\frac{\alpha i}{2(\alpha-1)}\left[\frac{\left( \frac{t+1}{t-1} \right)^{\alpha-1}-1}{(t+1)\left( \frac{t+1}{t-1} \right)^{\alpha-1}-(t-1)}\right].
\end{equation}
This will be useful for evaluating the contour integrals in Section \ref{sec:RenyiInts}.
We note that this last expression is also an odd function.

\section{A Convenient Basis for Polynomial Two-Vectors}\label{polybasis}
Let $\mathbb{C}[x_1,\dots,x_d]$ the vector space of $d$-variate complex polynomials. In this appendix, we will define a basis for the Cartesian product of this space with itself, denoted $\mathbb{C}[x_1,\dots,x_d]^2$, which is useful in the above work. 
Let $\eta$ encompass all $d-1$ spherical harmonic indices in $d$ dimensions. We begin by defining $G_\eta\in \mathbb{C}[x_1,\dots,x_d]^2$ by \begin{equation}\label{eq:GdefinitionApp}
G_\eta(\vectorbold x)=\begin{bmatrix}
    0\\ F^d_\eta(\bm{x}),
\end{bmatrix}
\end{equation}
where $F^d_{\eta}$ denotes the polynomials described in Appendix \ref{sec:harmPolys}.
Additionally, recall the definition of the linear map $J$ (c.f. Equation \ref{eq:J}), in this case restricted to $\mathbb{C}[x_1,\dots,x_d]^2$, 
\begin{equation}
    J\begin{bmatrix}\phi_1\\\phi_2\end{bmatrix}=\begin{bmatrix}
    \phi_2\\-\phi_1
\end{bmatrix}.
\end{equation}
Next, given the polynomial $p(\vectorbold{x})=1-|\vectorbold{x}|^2$, we define the linear map $H$ on $\mathbb{C}[x_1,\dots,x_d]^2$ as,
\begin{equation}\label{eq:HdefinitionApp}
H\begin{bmatrix}\phi_1\\\phi_2\end{bmatrix}=\frac{1}{2}\begin{bmatrix}
    \nabla\cdot(p\nabla\phi_2)-(d-1)\phi_2\\p\phi_1
\end{bmatrix}.
\end{equation}
We note that this linear map is $\frac{1}{\pi}\arccot(J\sigma_0)$, where $\sigma_0$ is the correlation matrix of the vacuum on the ball, which we can deduce by comparison with the modular Hamiltonian for a scalar vacuum. \cite{firstPaper,Casini_2010}
This operator respects spherical symmetry (i.e. it is equivariant with respect to rotation of the input). This  means that if $\phi_1$ and $\phi_2$ has the form $\phi(r,\bm{\theta})= f(r)Y_\eta(\bm{\theta})$, then the components of $H\begin{bmatrix}\phi_1\\\phi_2\end{bmatrix}$ have that same form. Use these operators to define the following set of polynomial 2-vectors:
\begin{equation}
     \mathcal{B}= \left\{JH^jG_\eta \bigg| 
 j \in \mathbb{Z}_{\geq 0}, \text{ }\eta \text{ ranges over all spherical harmonic indices}\right\}
\end{equation}
I claim that $\mathcal{B}$ is a basis for $\mathbb{C}[x_1,\dots,x_d]^2$. To see this, note that applying $H$ swaps the nonzero component. So, if $\mathcal{B}_{\text{odd}}$ and $\mathcal{B}_{\text{even}}$ denote the subsets of $\mathcal{B}$ where $j$ is odd and even respectively, then elements of $\mathcal{B}_{\text{odd}}$ only have nonzero top components and elements of $\mathcal{B}_{\text{even}}$ only have nonzero bottom components. In addition, applying $H$ increases the degree of the polynomial by two if the nonzero component is on the top but doesn't change the degree if the nonzero component on the bottom. As a consequence of the above facts and the fact that $H$ is spherically symmetric, the nonzero components of $\mathcal{B}_{\text{odd}}$ and $\mathcal{B}_{\text{even}}$ must each be related by a triangular system of equations to the set of polynomials: \begin{equation}
     \left\{r^{2j} F^d_\eta(\bm{x}) \bigg| 
 j \in \mathbb{Z}_{\geq 0}, \text{ }\eta \text{ ranges over all spherical harmonic indices}\right\}
\end{equation}
It is well known that this set is a basis for $\mathbb{C}[x_1,\dots,x_d]$\cite{axleraramey_1995,firstPaper} So, the nonzero components of $\mathcal{B}_{\text{odd}}$ and $\mathcal{B}_{\text{even}}$ must be a basis $\mathbb{C}[x_1,\dots,x_d]$. Therefore, $\mathcal{B}$ is a basis for $\mathbb{C}[x_1,\dots,x_d]^2$.

The choice to construct this basis out of the spherical harmonics allows one to take advantage of spherical or cylindrical symmetry in the system. It is worth noting that spherical harmonics contain many more symmetry properties than are useful in this problem. One unconcerned with taking full advantage of symmetry can use the same overall strategy presented in this work but replace the $F^d_\eta$ with fixed $\ell$ with any orthonormal (with respect to integration over the unit sphere) system of harmonic, homogeneous polynomials of degree $\ell$. It may be easier to calculate such a basis directly than rely the existing formulae for spherical harmonics present.

\section{Extending Spherical Harmonics to Harmonic Polynomials in Low Dimensions}\label{sec:harmPolys}
In arbitrary dimensions, given a spherical harmonic $Y_\eta$ with indices $\eta$ and highest index $\ell$, the function $F^d_\eta(\bm{x})\coloneq r^{|\ell|}Y_\eta(\bm{\theta})$ is a harmonic (i.e. satisfying Laplace's equation) polynomial in Cartesian coordinates. \cite{frye2012spherical} In this appendix, I give recurrence relations to compute these polynomials in any dimension.
In two dimensions, the spherical harmonics are labeled by an integer $m$. The corresponding harmonic polynomial is, 
\begin{equation}\label{eq:2DHarmPoly}
    F^{2}_{\pm m}(x_1,x_2)=\frac{1}{\sqrt{2\pi}}(x_1\pm ix_2)^m,  
\end{equation}
where we assume $m>0$.

For dimensions $d>2$, the spherical harmonics are labeled by $d-2$ nonnegative integers $\ell_0,\ell_1,\dots,\ell_{d-3}$ and an integer $m$ satisfying $|m|\leq \ell_{d-3}\leq  \dots \leq \ell_1 \leq \ell_{0} $. Define the Gegenbauer polynomials $C_\ell^{a}$ in terms of a generating function,
\begin{equation}\label{eq:Gegenbauer}
   \frac{1}{(1-2xt+t^2)^a}=\sum_{\ell=0}^\infty C_{\ell}^\alpha(x)t^\ell.
\end{equation}  The corresponding harmonic polynomials satisfy a recurrence relation using  Gegenbauer polynomials related them to lower dimensional harmonic polynomials. \cite{Meremianin_2009}
\begin{equation}\label{eq:nDHarmPoly}
    F^{d}_{\ell_0,\ell_1,\dots,\ell_{d-3},m}(x_1,\dots,x_{d-1},x_d)=2^{\gamma_1}\Gamma(\gamma_{1}) \sqrt{ \frac{  \gamma_0 (\gamma_0-\gamma_1)!}{2\pi (\gamma_0+\gamma_1-1)!}}
 r^{\gamma_0-\gamma_1}C^{\gamma_1}_{\gamma_0-\gamma_1}(\frac{x_d}{r})F^{d-1}_{\ell_1,\dots,\ell_{d-3},m}(x_1,\dots,x_{d-1}).  
\end{equation}
Here we define $\gamma_0=\frac{d-2}{2}+\ell_0$ and $\gamma_1=\frac{d-2}{2}+\ell_1$ for $d>3$. For $d=3$, we instead say $\gamma_1=\frac{d-2}{2}+|m|$. As well, $r$ denotes the radial coordinate with $r^2 = \sum_{i=1}^dx_i^2$. Since only even power of $r$ appear after cancellation, this is indeed a polynomial in the $x_i$. This formula can be used to recursively used to construct our harmonic polynomials for any dimension.

 \section{Integrals}\label{sec:integrals}
In this appendix, I compute three integral families needed in the main text, with each family covered in a sub-appendix. 
 \subsection{Computing the Matrix Elements}\label{sec:matrixElemsApp}
I will calculate, \begin{equation}
     M_{d,\ell,j,0}(z)=\frac{i}{\Gamma(\gamma)^2} \int_{-\infty}^{\infty}\diff u \text{ } \frac{(iu)^j\csch(\pi u)|\Gamma(\gamma+iu)|^2}{z+i\coth(\pi u)}.
\end{equation} I will only explicitly calculate this for integral $z \in \{z=ti\mid -1<t<1\}$, the analytic continuation to all $z\notin \{z=ti\mid |t|\geq 1\}$ will be obvious. We begin with the reparameterization $z = -i\tanh(\pi y)$,
\begin{equation}
    M_{d,\ell,j,0}(-i\tanh(\pi y)) = \frac{1}{\Gamma(\gamma)^2} \int_{-\infty}^{\infty}\diff u \text{ } \frac{(iu)^j\csch(\pi u)|\Gamma(\gamma+iu)|^2}{\coth(\pi u)-\tanh(\pi y)}.
\end{equation}
We can rewrite this using a hyperbolic identity for $\coth(a)-\tanh(b)$ and the identity $\frac{\pi}{\cosh(\pi x)}=|\Gamma(\frac{1}{2}+ix)|^2$ to get, 
\begin{equation}
     M_{d,\ell,j,0}(-i\tanh(\pi y))=\frac{\cosh(\pi y)}{\pi\Gamma(\gamma)^2} \int_{-\infty}^{\infty}\diff u \text{ } (iu)^j|\Gamma(\frac{1}{2}+i(u-y))|^2|\Gamma(\gamma+iu)|^2.
\end{equation}
Now we can use the binomial identity,
\begin{equation}
    (iu)^j = \sum_{m=0}^j{j \choose m} (iu+\gamma)^m(-\gamma)^{j-m},
\end{equation}
as well as the definition of the Stirling numbers of the second kind $\stirling{m}{k}$,
\begin{equation}
    x^m = \sum_{k=0}^m(-1)^{m-k}\stirling{m}{k} \frac{\Gamma(x+n)}{\Gamma(x)},
\end{equation}
to rewrite $M_{d,\ell,j,0}$ as a sum of integrals:
\begin{align}
    \begin{split}
         M_{d,\ell,j,0}(-i\tanh(\pi y))=&\frac{\cosh(\pi y)}{\pi\Gamma(\gamma)^2}\sum_{k=0}^j\Biggl[\left(\sum_{m=0}^{j-k}(-1)^{j-k} {j \choose m} \stirling{j-m}{k}\gamma^m\right)
         \\&\int_{-\infty}^{\infty}\diff u \text{ } |\Gamma(\frac{1}{2}+i(u-y))|^2\Gamma(\gamma+k+iu)\Gamma(\gamma-iu)\Biggr]
    \end{split}
\end{align}
These integrals can now be computed with Barnes' beta integral \cite{barnes},
\begin{equation}
M_{d,\ell,j,0}(-i\tanh(\pi y))=\frac{2\cosh(\pi y)}{\Gamma(\gamma)^2}\sum_{k=0}^j\frac{\sum_{m=0}^{j-k}(-1)^{j-k} {j \choose m} \stirling{j-m}{k}\gamma^m}{2\gamma+k} \Gamma\left(\gamma+k+\frac{1}{2}+iy\right) \Gamma\left(\gamma+\frac{1}{2}-iy\right).
\end{equation}
We substitute back in $y = i\frac{\arctan{z}}{\pi}$ for $z \in \{z=yi\mid -1<y<1\}$. If $\gamma$ is an integer, $M_{d,\ell,j,0}(z)$ has the form, \begin{equation}
    M_{d,\ell,j,0}(z) =  \frac{\pi}{\gamma\Gamma(\gamma)^2}P_{\gamma,j}(\frac{\arctan(z)}{\pi}), 
\end{equation} where $P_{\gamma,j}$ is a rational polynomial satisfying $P_{\gamma,j}(\frac{1}{2})=P_{\gamma,j}(-\frac{1}{2})=0$. If $\gamma$ is a half-integer, $M_{d,\ell,j,0}(z)$ has the form, \begin{equation}
    M_{d,\ell,j,0}(z) =  \frac{\pi}{\gamma\Gamma(\gamma)^2z}P_{\gamma,j}(\frac{\arctan(z)}{\pi}), 
\end{equation} where $P_{\gamma,j}$ is a rational polynomial satisfying $P_{\gamma,j}(0)=0$. We can now comfortably analytically extend these results to everywhere on the domain of the principal branch of $\arctan$. This is exactly the set of all $z\notin\mathrm{Spec}(J\sigma_0)$. 
Concretely, we can write $ P_{\gamma,j}$ as,
\begin{equation}
    P_{\gamma,j}(x) = P_{\gamma,0}(x)\sum_{k=0}^j\frac{2\gamma}{2\gamma+k}\left[\sum_{m=0}^{j-k}(-1)^{j-k} {j \choose m} \stirling{j-m}{k}\gamma^m\right]\left(\gamma+\frac{1}{2}-x\right)^{(k)}
    \end{equation}
where, $t^{(k)}=\frac{\Gamma(t+k)}{\Gamma(t)}=\prod_{n=0}^{k-1}
(t+n)$. If $\gamma$ is an integer, $ P_{\gamma,0}$ can be written as,
\begin{equation}
    P_{\gamma,0}(x) =  \prod_{k=1}^\gamma ((k-\frac{1}{2})^2-x^2),
\end{equation}
and $\gamma$ is a half integer $ P_{\gamma,0}$ is,
\begin{equation}
    P_{\gamma,0}(x) = x\prod_{k=1}^{\gamma-\frac{1}{2}} (k^2-x^2).
\end{equation}
  \subsection{The Radial Kernel for Spherically Symmetric Systems}\label{sec:radialKernel}
We calculate the integral found in Section \ref{symmetry} for $d\geq 3$\footnote{The $d=2$ case is not necessary for this work.},
\begin{equation}
    W^{q}_\ell(r,s)Y_{\eta_\ell}(\bm{\alpha}) = \int_{\partial \Omega} \diff^{d-1}\bm{\theta}\text{ } |\vectorbold{x}-\vectorbold{y}|^{2q} Y_{\eta_\ell}(\bm{\theta}),
\end{equation}
where $\vectorbold{x}=(r,\bm{\theta})$ and $\vectorbold{y}=(s,\bm{\alpha})$ in spherical coordinates and $Y_{\eta_\ell}$ is the spherical harmonic with first index $\ell$ and all other indices zero. Set the Cartesian form of $\vectorbold{y}$ to be $(0,\dots,0,s)$ so that it lies on the north pole. Furthermore, the spherical harmonic $Y_{\eta_\ell}$ is proportional to a Gegenbauer polynomial \cite{Meremianin_2009},
\begin{equation}
    Y_{\eta_\ell}(\bm{\theta}) \propto C_\ell^{\frac{d}{2}-1}(\cos(\theta_1)),
\end{equation}
where the Gegenbauer polynomials are defined in Equation \ref{eq:Gegenbauer}). We can now write the integral as,
\begin{equation}
    W^{q}_\ell(r,s)= \frac{1}{C_\ell^{\frac{d}{2}-1}(1)}\int_{\partial \Omega} \diff^{d-1}\bm{\theta}\text{ } (r^2+s^2-2rs\cos(\theta_1))^{q} C_\ell^{\frac{d}{2}-1}(\cos(\theta_1)).
\end{equation}
The integrals over $\theta_2,\dots,\theta_{d-1}$ evaluate to the surface area of the $(d-2)$-sphere. In addition, we rewrite the integral over $\theta_1\in [0,\pi]$ with the substitution $u=\cos(\theta_1)$,
\begin{equation}
    W^{q}_\ell(r,s)= \frac{2\pi^{\frac{d}{2}-1}}{\Gamma(\frac{d-1}{2})C_\ell^{\frac{d}{2}-1}(1)}\int_{-1}^{1} \diff u\text{ } (r^2+s^2-2rsu)^{q} (1-u^2)^{\frac{d-3}{2}} C_\ell^{\frac{d}{2}-1}(u).
\end{equation}
We expand the using the binomial theorem,
\begin{equation}
    W^{q}_\ell(r,s)=\frac{2\pi^{\frac{d}{2}-1}}{\Gamma(\frac{d-1}{2})C_\ell^{\frac{d}{2}-1}(1)}\sum_{k=0}^{q}{q\choose k} (r^2+s^2)^{q-k}(-2rs)^{k} \int_{-1}^1 \diff u\text{ }u^k(1-u^2)^{\frac{d-3}{2}}C_\ell^{\frac{d}{2}-1}(u).
\end{equation}
Note that the resulting integrals are zero if $k$ and $\ell$ have the opposite parity because the integrand is odd and is zero if $k<\ell$ due to the orthogonality of the $C_\ell^{\frac{d}{2}-1}$. The remaining integrals can be solved using 7.311.2 in Gradstein and Ryzhik. \cite{2014776} We can also compute $C_\ell^{\frac{d}{2}-1}(1)$ using the generating function in Equation \ref{eq:Gegenbauer}. The final result for $W^{q}_\ell$ is
\begin{equation}
   W^{q}_\ell(r,s)=2q!\pi^{\frac{d}{2}} \sum_{\rho=0}^{\lfloor\frac{q-\ell}{2}\rfloor} \frac{(-rs)^{2\rho+\ell}(r^2+s^2)^{q-2\rho-\ell}}{\rho!(q-\ell-2\rho)!\Gamma(\frac{d}{2}+\rho+\ell)}.
\end{equation}

\subsection{Computing
the Contour Integral for Rényi Entropy}\label{sec:RenyiInts} 
To compute the Rényi entropy $S_\alpha$, we must consider the general family of contour integrals given by, 
\begin{equation}\label{eq:contourForm0}
    T_{n,W}(\alpha)=\frac{1}{2\pi}\int_{\xi} \diff z\text{ } \frac{\alpha}{2(\alpha-1)}\frac{W(\arctan(z))}{z^n}\left[\frac{e^{2(\alpha-1) i \arccot(z)}-1}{(1-iz)e^{2(\alpha-1) i \arccot(z)}+(1+iz)}\right]
\end{equation}
where $n$ is a nonnegative integer and $0\leq a\leq \infty$. For $a=0,1,\infty$, we define this integral by continuity. Here, \begin{equation}
     W(x)=\sum_{j=n}^{\deg(W)} w_j x^j
 \end{equation} is a polynomial with a degree $n$ root at the origin and $\xi$ is the closed contour shown in Figure \ref{fig:double_keyhole}. We also note that if $n=0$, $W$ satisfies $W(\frac{\pi}{2})=W(-\frac{\pi}{2})=0$. As a function, $W$ has the same parity as $x^n$. 

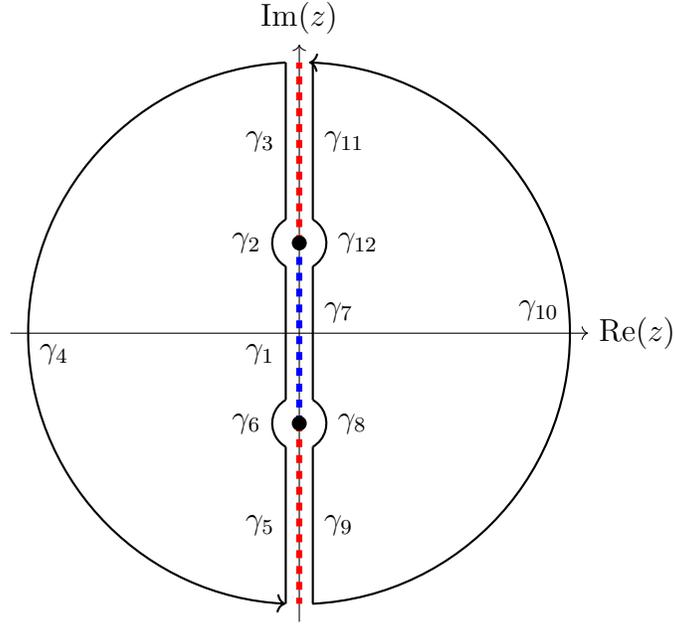
\begin{figure}[htbp]
    \centering
    \begin{tikzpicture}[scale=1.2]
        \draw[ ->] (-3.2,0) -- (3.2,0) node[right] {$\Re(z)$};
        \draw[ ->] (0,-3.2) -- (0,3.2) node[above] {$\Im(z)$};
        
        \draw[thick] (-0.15, 1.25980) arc[start angle=-240, end angle=-120, radius=0.3]
        node[left,midway] {$\gamma_2$}; 
        \draw[thick] (0.15,0.7402) arc[start            angle=-60, end angle=60, radius=0.3]
        node[right,midway] {$\gamma_{12}$};

        \draw[thick] (-0.15, -0.7402) arc[start angle=-240, end angle=-120, radius=0.3]
        node[left,midway] {$\gamma_6$}; 
        \draw[thick] (0.15,-1.25980) arc[start            angle=-60, end angle=60, radius=0.3]
        node[right,midway] {$\gamma_{8}$};

        \draw[thick] (-0.15, 1.25980)-- node[left,midway] {$\gamma_3$}(-0.15, 3);
        \draw[thick] (0.15, 1.25980)-- node[right,midway] {$\gamma_{11}$} (0.15, 3);
        \draw[thick] (-0.15, -1.25980)-- node[left,midway] {$\gamma_5$} (-0.15, -3);
        \draw[thick] (0.15, -1.25980)-- node[right,midway] {$\gamma_9$}(0.15, -3);
        \draw[thick] (-0.15, -0.7402)-- node[below left,midway] {$\gamma_1$}(-0.15, 0.7402);
        \draw[thick] (0.15, -0.7402)-- node[above right,midway] {$\gamma_7$}(0.15, 0.7402);
        
        \draw[thick,->] (-0.15, 3) arc[start angle=92.862405, end angle=267.1375, radius=3.00374765918] node[below right,midway] {$\gamma_4$};
        \draw[thick,->] (0.145, -3) arc[start angle=-87.1375, end angle=87.862405, radius=3.00374765918] node[above left,midway] {$\gamma_{10}$};
        
        \draw[line width=0.8mm, dashed, red] (0,-1) -- (0,-3);
        \draw[line width=0.8mm, dashed, red] (0,1) -- (0,3);
        \draw[line width=0.8mm, dashed, blue] (0,-1) -- (0,1);
        
         \node at (0,1)[circle,fill,inner sep=2pt]{};
         \node at (0,-1)[circle,fill,inner sep=2pt]{};
    \end{tikzpicture}
    \caption{A double keyhole contour which avoids the branch cuts of $\arctan$ (shown in red) and $\arccot$ (shown in blue). I label each smooth component with a $\gamma_i$. I am focused on the limiting case when the outer circles' radii goes to infinity, the inner circles' radii goes to zero, and the lines approach the dotted lines. Integration is taken counter clockwise.}
    \label{fig:double_keyhole2}
\end{figure}

It will be convenient to place the disconnected contour in Figure \ref{fig:double_keyhole} inside of the closed contour shown in Figure \ref{fig:double_keyhole2}, letting us write the contour integral as 
\begin{equation}\label{eq:contourForm}
    T_{n,W}(\alpha)=\frac{1}{2\pi}\int_{\substack{\gamma_3+\gamma_{11}+\\\gamma_5+\gamma_9}}  \diff z\text{ } \frac{\alpha}{2(\alpha-1)}\frac{W(\arctan(z))}{z^n}\left[\frac{e^{2(\alpha-1) i \arccot(z)}-1}{(1-iz)e^{2(\alpha-1) i \arccot(z)}+(1+iz)}\right],
\end{equation}
where the $\gamma_i$ are shown in Figure \ref{fig:double_keyhole2}.
We can also explicitly parameterize the contour integrals to get a real integral,
\begin{equation}\label{eq:differenceForm}
    T_{n,W}(\alpha)= \frac{\alpha}{4(1-\alpha)\pi}\int_{(-\infty,-1)\cup (1,\infty)} \diff{t}\text{ }\frac{V(\arccoth(t))}{t^n} \left[\frac{\left( \frac{t+1}{t-1} \right)^{\alpha-1}-1}{(t+1)\left( \frac{t+1}{t-1} \right)^{\alpha-1}-(t-1)}\right],
\end{equation}
where $V(x)$ is shorthand for $V(x)=i^{-n-1}(W(ix+\frac{\pi}{2})-W(ix-\frac{\pi}{2}))$. The binomial theorem allows us to write $V$ explicitly as
\begin{equation}\label{eq:Vform}
   V(x)= 2\sum_{j=n}^{\deg(W)}w_j\sum_{k=0}^{\lceil \frac{j}{2}\rceil-1}{j \choose 2k+1} \left(\frac{\pi}{2}\right)^{2k+1}(-1)^{\frac{n-j}{2}+k}x^{j-1-2k}
\end{equation}
We solve this if $n=0$, which happens if the spatial dimension of the field theory is odd in Appendix \ref{sec:oddDimCalc}. But in general, this is difficult to compute analytically if $n\geq 1$, which happens if the spatial dimension of the field theory is even. However, we look at the important special cases for $n\geq 1$ in this appendix.

\begin{figure}[htbp]
    \centering
    \begin{tikzpicture}[scale=1.2]
        \draw[ ->] (-3.2,0) -- (3.2,0) node[right] {$\Re(z)$};
        \draw[ -] (0,-3.2) -- (0,-0.7);
        \draw[ -] (0,-0.4) -- (0,0.4);
         \draw[ ->] (0,0.7) -- (0,3.2) node[above] {$\Im(z)$};
        
        \draw[thick,->] (-0.15,-0.7402) arc[start  angle=120, end angle=420, radius=0.3]
        node[below left,midway] {$\gamma_{6}$};

        \draw[thick,->] (0.15, 0.7402) arc[start angle=-60, end angle=240, radius=0.3]
        node[above right,midway] {$\gamma_2$}; 

        \draw[thick] (-0.15,0.2598) arc[start angle=120, end angle=240, radius=0.3]
        node[above left,midway] {$\gamma_{4}$};
        \draw[thick] (0.15,-0.2598) arc[start   angle=-60, end angle=60, radius=0.3]
        node[below  right, midway] {$\gamma_{8}$};

         \draw[thick] ( 0.15,-0.7402)-- node[right,midway] {$\gamma_7$}( 0.15,-0.2598);
        \draw[thick] (-0.15,-0.7402)-- node[left,midway] {$\gamma_5$}(-0.15,-0.2598);
         \draw[thick] (0.15, 0.2598)-- node[right,midway] {$\gamma_1$}(0.15, 0.7402);
        \draw[thick] (-0.15,0.2598)-- node[left,midway] {$\gamma_3$}(-0.15,0.7402);

        \draw[line width=0.8mm, dashed, red] (0,-1) -- (0,1);
        
         \node at (0,1)[circle,fill,inner sep=2pt]{};
         \node at (0,-1)[circle,fill,inner sep=2pt]{};
         \node at (0,0)[circle,fill,inner sep=2pt]{};
    \end{tikzpicture}
    \caption{A double keyhole contour enclosing the portion of the real line with magnitude less than one, deformed around a pole at the origin. The branch cut is shown with a red dotted line. I am focused on the limiting case when the circles' radii goes to zero, and the lines approach the branch cut.  Integration is taken counter clockwise.}
    \label{fig:double_keyhole3}
\end{figure}
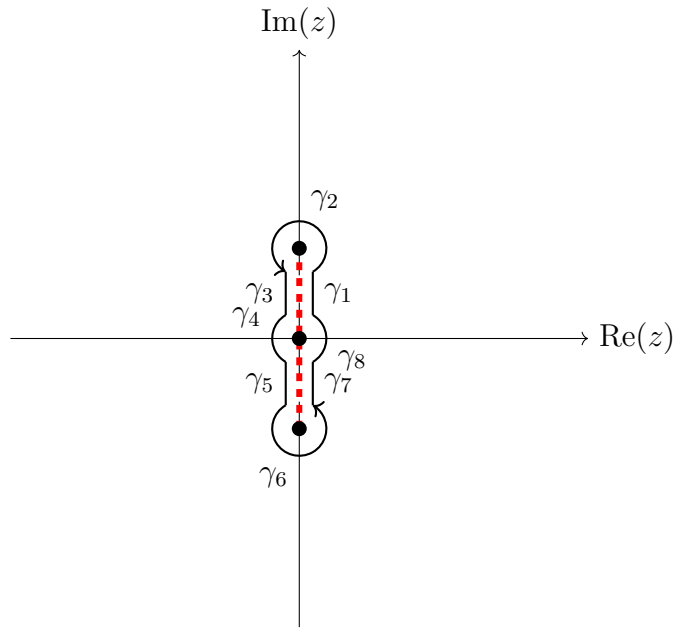

\subsubsection{Rényi Entropy in odd dimensions}\label{sec:oddDimCalc}
We begin by looking at the case of $n=0$, which is the only case needed to compute Rényi entropy in odd dimensions for all $\alpha$. We start with Equation \ref{eq:differenceForm}. Note that the integrand is even, meaning we only need to consider the positive portion of the integration region,
\begin{equation}
     T_{0,W}(\alpha)= \frac{\alpha}{2(1-\alpha)\pi}\int_1^\infty \diff{t}\text{ } V(\arccoth(t))\left[\frac{\left( \frac{t+1}{t-1} \right)^{\alpha-1}-1}{(t+1)\left( \frac{t+1}{t-1} \right)^{\alpha-1}-(t-1)}\right],
\end{equation}
where $V(x)$ is given in Equation \ref{eq:Vform}. Note that $V$ is an odd function. We now make the substitution $u=\frac{t+1}{t-1}$ and ,
\begin{equation}
     T_{0,W}(\alpha)= \frac{\alpha}{2(1-\alpha)\pi}\int_0^1\diff{t}\text{ } V\left(\frac{\ln(u)}{2}\right)\left[\frac{1}{u-1}-\frac{u^{\alpha-1}}{u^{\alpha}-1}\right].
\end{equation}
If we make the substitution $y=u^\alpha$ on only the right term, we get,
\begin{equation}
    T_{0,W}(\alpha)= \frac{\alpha}{2(1-\alpha)\pi}\int_0^1\diff{t}\text{ } \frac{V\left(\frac{\ln(u)}{2}\right)-\frac{1}{\alpha}V\left(\frac{\ln(u)}{2\alpha}\right)}{u-1}.
\end{equation}
Now, we substitute in Equation \ref{eq:Vform},

\begin{equation}
    T_{0,W}(\alpha)= \frac{\alpha}{(1-\alpha)\pi}\sum_{j=0}^{\deg(W)}w_j\sum_{k=0}^{\frac{j}{2}-1}\frac{{j \choose 2k+1} \left(\frac{\pi}{2}\right)^{2k+1}(-1)^{\frac{j}{2}+k}}{2^{j-1-2k}} \int_0^1\diff{t}\text{ } \frac{(1-\alpha^{2k-j})\ln(u)^{j-1-2k}}{u-1}.   
\end{equation}
Using 4.271.3 in Gradstein and Ryzhik \cite{2014776},  we get

\begin{equation}
    T_{0,W}(\alpha)= \frac{\alpha}{1-\alpha}\sum_{j=0}^{\deg(W)}\frac{w_j\pi^j}{j+1}\sum_{k=0}^{\frac{j}{2}-1}\left[\frac{{j+1 \choose 2k+1}}{2^{2k+1}} (\alpha^{2k-j}-1)B_{j-2k}\right].   
\end{equation}
Using the definition of the Bernoulli polynomials (see Appendix \ref{sec:bernoulli}), we can write this as
\begin{equation}
    T_{0,W}(\alpha)= \frac{1}{2}W(\frac{\pi}{2})+\frac{1}{1-\alpha}\sum_{j=0}^{\deg(W)}\frac{w_j}{j+1}\left(\frac{\pi}{\alpha}\right)^j\left(B_{j+1}\left(\frac{\alpha}{2}\right)-B_{j+1}\left(\frac{1}{2}\right)\right).
\end{equation}
We can further simplify because $W(\frac{\pi}{2})=0$ and, since $j$ is even, $B_{j+1}\left(\frac{1}{2}\right)=0$
\begin{equation}
    T_{0,W}(a)= \sum_{j=0}^{\deg(W)}w_j\left(\frac{\pi}{\alpha}\right)^j 
    \left[\frac{B_{j+1}(\frac{\alpha}{2})}{(j+1)(1-\alpha)}\right].
\end{equation}
In particular, for Von Neumann entropy we substitute $\alpha=1$,
\begin{equation}
    T_{0,W}(1)= \frac{1}{2}\sum_{j=0}^{\deg(W)}w_j\pi^j 
    (1-2^{1-j})B_j.
\end{equation}

\subsubsection{Von Neumann Entropy in Even Dimensions }\label{sec:a=1Even}
For even spatial dimensions which must have $n\geq 1$, the most important case is when $\alpha=1$, which is needed to compute Von Neumann entropy.  For Von Neumann entropy, Equation \ref{eq:contourForm} becomes 
\begin{equation}
      T_{n,W}(1)=\frac{i}{4\pi }\int_{\substack{\gamma_3+\gamma_{11}+\\\gamma_5+\gamma_9}} \diff z\text{ } \frac{\arccot(z)W(\arctan(z))}{z^n},
\end{equation}
where the $\gamma_i$ are portions of the closed contour shown in Figure \ref{fig:double_keyhole2}. Note that since the integrand is analytic inside the contour, the entire contour integral evaluates to zero by Cauchy's integral theorem. In addition, all circular contours tend to zero by standard ML lemma arguments\footnote{In particular, the fact that $W(\frac{\pi}{2})=W(-\frac{\pi}{2})=0$ if $n=0$ is what allows the contour along the outer circles to vanish.}.
Therefore,
\begin{equation}
      T_{n,W}(1)=\frac{1}{4\pi i}\int_{\gamma_1+\gamma_{7}} \diff z\text{ }\frac{\arccot(z)W(\arctan(z))}{z^n},
\end{equation}
We can parameterize this contour integral as,
\begin{equation}
      T_{n,W}(1)=-\frac{(-i)^{n}}{4}\int_{-1}^1 \diff t\text{ }\frac{W(i\arctanh(t))}{t^n},
\end{equation}
This integral is straightforward enough to plug into a symbolic integrator for any specific $W$ and $n$. Regardless, I will compute the general form here. In particular, the final answer will be a rational polynomial in $\pi$. 

Let $U(x)$ be a polynomial satisfying $U(-\frac{\pi}{2}-x)-U(\frac{\pi}{2}-x)= W(x)$. Concretely, we can write such a $U$ in terms of Bernoulli polynomials (see Section \ref{sec:bernoulli}) as,
\begin{equation}\label{eq:Udfn}
    U(x)=\sum_{j=n}^{\deg(W)} \pi^j \frac{w_j}{j+1}B_{j+1}\left(-\frac{x}{\pi}+\frac{1}{2}\right).
\end{equation}
We can write $T_{n,W}(1)$ as the following contour integral, 

\begin{equation}
      T_{n,W}(1)=\frac{1}{4i}\int_{\substack{\gamma_1+\gamma_{3}+\\\gamma_5+\gamma_7}} \diff z\text{ }\frac{U(\arccot(z))}{z^n},
\end{equation}
around portions of the contour shown in Figure \ref{fig:double_keyhole3}. Noting that the integrals around $\gamma_6$ and $\gamma_2$ in Figure \ref{fig:double_keyhole3} don't contribute (again due to ML Lemma arguments), we can write $T_{n,W}(1)$ as
\begin{equation}\label{eq:EntropyEvenDimContour2}
      T_{n,W}(1)=\frac{i}{4}\int_{\gamma_4+\gamma_{8}} \diff z\text{ }\frac{U(\arccot(z))}{z^n}-\frac{\pi}{2}\Res(\frac{U(\arccot(z))}{z^n},\infty).
\end{equation}
 To compute the integrals over $\gamma_4$ and $\gamma_{8}$, observe that if $\Re(z)>0$ then $\arccot(z)=-\arctan(z)+\frac{\pi}{2}$ and if $\Re(z)<0$ then $\arccot(z)=-\arctan(z)-\frac{\pi}{2}$, where $\arctan(z)$ has a branch cut on the portion of the imaginary line with magnitude greater than $1$. So, we simplify the first term as,

 \begin{align}
     \begin{split}
        \frac{i}{4}\int_{\gamma_4+\gamma_{8}} \diff z\text{ }\frac{U(\arccot(z))}{z^n}&=-\frac{\pi}{4} \Res(\frac{U(-\arctan(z)+\frac{\pi}{2})+U(-\arctan(z)-\frac{\pi}{2})}{z^n},0),\\&=-\frac{\pi}{2}  \Res(\frac{U(-\arctan(z)+\frac{\pi}{2})}{z^n},0)+\frac{\pi}{2} \Res(\frac{W(-\arctan(z))}{z^n},0),\\&=-\frac{\pi}{2}\mathrm{Taylor}(U(-\arctan(z)+\frac{\pi}{2}),n-1,0),\\
         &=-\frac{\pi}{2}\sum_{j=n}^{\deg(W)}\pi^j \frac{w_j}{j+1}\mathrm{Taylor}\left(B_{j+1}\left(\frac{\arctan(z)}{\pi}\right),n-1,0\right),\\&=-\sum_{j=n}^{\deg(W)} \frac{w_j}{2(j+1)}\left[\sum_{k=j-n+2}^j{j+1\choose k}B_{k} \pi^k\mathrm{Taylor}(\arctan(z)^{j+1-k},n-1,0)\right].
     \end{split}
 \end{align}
Here we denote by $\mathrm{Taylor}(f(z),k,z_0)$ the $k$-th coefficient of the Taylor series of $f(z)$ at $z_0$. Note that to get to the second line, we use the property $U(\frac{\pi}{2}+x)-U(\frac{\pi}{2}-x)= W(x)$ and to get to the third line we note that $W(x)$ and therefore $W(-\arctan(z))$ has a zero of order $n$ or larger at zero, meaning the second residue is zero. We then use Equation \ref{eq:Udfn} and the expression for Bernoulli polynomials in Appendix \ref{sec:bernoulli}

On the other hand, the residue at infinity in Equation \ref{eq:EntropyEvenDimContour2} is zero for $n>1$. For $n=1$ it is,
\begin{align}
     \begin{split}
        \Res(\frac{U(\arccot(z))}{z},\infty) &= -\Res(\frac{U(\arctan(z))}{z},0)\\
        &=-U(0)\\
        &=-\sum_{j=n}^{\deg(W)} \pi^j \frac{w_j}{j+1}B_{j+1}\left(\frac{1}{2}\right)\\
        &=-\sum_{j=n}^{\deg(W)} \pi^j \frac{w_j}{j+1}(2^{-j}-1)B_{j+1}.
        \end{split}
 \end{align}
So we can write the contour integral for $n=1$ as,
\begin{equation}
    T_{1,W}(1)=\sum_{j=n}^{\deg(W)} \pi^{j+1} \frac{w_j}{j+1}(2^{-j-1}-1)B_{j+1}
\end{equation}
For $n > 1$, we can compute the Taylor series coefficients explicitly to write the contour integral as, noting that $k-n$ must be even and nonnegative and that $B_{2k+1}=0$ for $k\geq 1$. 
\begin{equation}
    T_{n,W}(1)=-\sum_{j=n}^{\deg(W)} \frac{w_j}{2(j+1)}\left[\sum_{k=\frac{j-n}{2}+1}^{\lfloor\frac{j}{2}\rfloor}\pi^{2k}{j+1\choose 2k}B_{2k} \sum_{\substack{\sum_{l=1}^{j+1-2k} q_l = n-1\\q_l \text{ odd}}}\left(\prod_l\frac{1}{q_l}\right)\right].
\end{equation}

\subsubsection{Integer Rényi entropy in even dimensions}\label{sec:aInteger}
The third case is when $\alpha$ is a positive integer greater than one, which allows us to compute Rényi entropy for $\alpha=2,3,4,\dots$ for even dimensions. In this case, the integrand of equation \ref{eq:contourForm} becomes, 
\begin{equation}\label{eq:aInteger}
    T_{n,W}(\alpha)=\frac{\alpha}{4\pi(1-\alpha)i}\int_{\substack{\gamma_3+\gamma_{11}+\\\gamma_5+\gamma_9}} \diff z\text{ } \frac{W(\arctan(z))}{z^n}\left[\frac{\left( z+i \right)^{\alpha-1}-\left( z-i \right)^{\alpha-1}}{\left( z+i \right)^{\alpha}-\left( z-i \right)^{\alpha}}\right].
\end{equation}
Note that the integrand is meromorphic on the interior of the contour in Figure \ref{fig:double_keyhole2}. Moreover, thanks to ML Lemma arguments, the only component that contributes to this closed contour is the one in Equation \ref{eq:aInteger}\footnote{Technically, the contours at $\gamma_1$ and $\gamma_7$ would contribute if there was a pole on the imaginary line between them. But, the contribution would be $2\pi i$ times the residue of the pole. But the pole would also be excluded from the interior of the contour, meaning Equation \ref{eq:aIntegerfinalApp} is still correct.}.  The integrand will have $\alpha-1$ poles, all first-order, at the points $z_j=\cot(\frac{j\pi}{\alpha})$ for $1\leq j<\alpha$. So, we can write this integral as the sum of the residues at these points,

\begin{equation}\label{eq:aIntegerfinalApp}
     T_{n,W}(\alpha)=\frac{1}{2(1-\alpha)} \sum_{j=1}^{\alpha-1} \tan(\frac{j\pi}{\alpha})^n W(\frac{\pi}{2}-\frac{j\pi}{\alpha}).
\end{equation}
If $2j=\alpha$, the term is defined by continuity, since $W$ has an order $n$ root at zero.
\subsubsection{Min-Entropy in Even Dimensions}
Now, we look at $\alpha\to \infty$, which is often called the min-entropy, in even dimensions. One approach would be to take the limit of Equation \ref{eq:differenceForm} as $\alpha\to\infty$, which produces a convergent integral. But it will be simpler to instead take advantage of the work in Section \ref{sec:aInteger} for integer $\alpha$. Note that Equation \ref{eq:aIntegerfinalApp} is a Riemann sum. So, as $a\to \infty$, it becomes,
\begin{equation}
    T_{n,W}(\infty)=-\frac{1}{2\pi}\int_0^{\pi}\diff{x}\text{ }\tan(x)^nW(\frac{\pi}{2}-x).
\end{equation}
Recall that $W$ has an order $n$ root at the origin, so the integrand is smooth. Make the substitution $u=\frac{\pi}{2}-x$,
\begin{equation}
    T_{n,W}(\infty)=-\frac{1}{2\pi}\int_{-\frac{\pi}{2}}^{\frac{\pi}{2}}\diff{x}\text{ }\cot(x)^nW(x).
\end{equation}
It suffices to compute the family of integrals:
\begin{equation}
    I(m,n)=-\frac{1}{2\pi}\int_{-\frac{\pi}{2}}^{\frac{\pi}{2}}\diff{x}\text{ }\cot(x)^nx^m.
\end{equation}
Note that $I(j,m)=0$ if $j-m$ is odd. Trivially, we get,
\begin{equation}
    I(2j,0)=-\frac{1}{2(2j+1)}\left(\frac{\pi}{2}\right)^{2j}.
\end{equation}
Let us also compute $I(2j+1,1)$. We begin by substituting $x=\pi t$ and using Equation \ref{eq:shiftBern},
\begin{equation}
    I(2j+1,1) = -\frac{\pi^{2j+1}}{2}\sum_{k=0}^j\frac{{2j+2 \choose 2k+1}}{(j+1)2^{2j-2k}}\int_0^{\frac{1}{2}}\diff{t}\text{ }B_{2k+1}(\frac{1}{2}-t)\cot(\pi t).
\end{equation}
Note that we have taken advantage of the evenness of the integrand and restricted ourselves to the positive half of the integration region. Substituting $u=\frac{1}{2}-x$ and using Theorem 1 in \cite{bernoulliInts}, we can write 
\begin{equation}
    I(2j+1,1) = \frac{(2j+1)!}{2^{2j+1}}\sum_{k=0}^j \frac{(-1)^k\pi^{2(j-k)}(4^{-k}-1)}{(2j-2k+1)!}\zeta(2k+1).
\end{equation}
Here $\zeta(s)=\sum_{j=1}^\infty\frac{1}{j^s}$ is the Riemann zeta function. With $I(2j,0)$ and $I(2j+1,1)$, we can compute all $I(m,n)$ with the following recurrence formula, which is easily proven with integration by parts,
\begin{equation}
    I(m,n) = (\frac{m}{n-1})I(m-1,n-1)-I(m,n-2).
\end{equation}

\subsubsection{Max-Entropy in Even Dimensions}
Lastly, we also look at when $\alpha=0$, often called the max-entropy, in even spatial dimensions. Taking the limit as $\alpha\to 0$ of Equation \ref{eq:differenceForm} gives,
\begin{equation}
T_{n,W}(0)=-\frac{1}{2\pi}\int_1^\infty \diff{t}\text{ }\frac{V(\arccoth(t))}{t^n} \left[\frac{1}{(t^2-1)\arccoth(t)}\right].
\end{equation}
This is always a divergent integral. So, we expect the terms of the Rényi entropy difference to diverge as $\alpha\to 0$.

\section{Bernoulli Polynomials and Numbers}\label{sec:bernoulli}

In this section, we define the Bernoulli polynomials, and discuss selected properties from \cite{abramowitz+stegun} which are important for computing the various integrals in Appendix \ref{sec:integrals}. Consider the following functional equation,
\begin{equation}\label{eq:BernFuncEq}
    F(x+1)-F(x)=nx^{n-1}.
\end{equation}
This functional equation has a solution in the polynomials up to an additive constant. The $n$th \textit{Bernoulli polynomial} $B_n(x)$ is defined as the solution to this equation with $\int_0^1\diff x B_n(x) = 0$, which fixes the additive constant. The exception is that $B_0(x)$ is simply equal to $1$. We also define the $n$-th \textit{Bernoulli number} $B_n$\footnote{Using the same symbol for the polynomial and number is awkward but it is standard notation.} as 
\begin{equation}
    B_n=B_n(0).
\end{equation} Note that $B_{2k+1}=0$ for $k>0$. The Bernoulli polynomials also satisfy $B_n(1)=(-1)^nB_n$, and $B_n(\frac{1}{2})=(2^{1-n}-1)B_n$. The Bernoulli polynomials satisfy the differential relation $B_{n}'(x)=nB_{n-1}(x)$. This relation allows us to entirely write the Bernoulli polynomials in terms of the Bernoulli numbers and binomial coefficients:
\begin{equation}
    B_n(x)=\sum_{k=0} {n\choose k}B_kx^{n-k}.
\end{equation}
Lastly, we have an identity which is not typically discussed but is useful for our purposes,
\begin{equation} \label{eq:shiftBern}
    x^{2j+1}=\sum_{k=0}^j\frac{{2j+2 \choose 2k+1}}{(j+1)2^{2j-2k+1}}B_{2k+1}(x+\frac{1}{2}),
\end{equation}
This can easily be checked with the defining properties of the Bernoulli polynomials by verifying both sides give the same result when shifted by $\frac{1}{2}$ then substituted into the functional equation in Equation \ref{eq:BernFuncEq} and also when integrated from $-\frac{1}{2}$ to $\frac{1}{2}$.

\section{The Expansions for Rényi Entropy}\label{sec:RenyiFull}
In this appendix, we present the expansions described in Sections \ref{sec:MI} and \ref{sec:Ent} for $\alpha \neq 1$.
\subsection{Rényi Mutual Information of Distant Balls}\label{sec:RenyMIFull}
\subsubsection{Two Dimensions}
For  $\alpha = 2$ , we have,
\begin{align}
    I_{2,2}(r, R) &= \frac{2}{\pi^2} \left(\frac{r}{R}\right)^2 + \left(\frac{4}{\pi^{2}}+\frac{4}{\pi^{4}}\right) \left(\frac{r}{R}\right)^4 +  \left(\frac{112}{9 \pi^{2}}+\frac{16}{\pi^{4}}+\frac{32}{3 \pi^{6}}\right) \left(\frac{r}{R}\right)^{6} + \mathcal{O}\left(\left(\frac{r}{R}\right)^{8}\right) \\
    &= 0.203 \left(\frac{r}{R}\right)^2 + 0.446 \left(\frac{r}{R}\right)^4 + 1.436 \left(\frac{r}{R}\right)^{6} + \mathcal{O}\left(\left(\frac{r}{R}\right)^{8}\right).
\end{align}
For  $\alpha = 3$ , we have,
\begin{align}
    I_{2,3}(r, R) &= \frac{1}{6} \left(\frac{r}{R}\right)^2 + \frac{10}{27} \left(\frac{r}{R}\right)^4 + \frac{10381}{8748} \left(\frac{r}{R}\right)^{6} + \mathcal{O}\left(\left(\frac{r}{R}\right)^{8}\right) \\
    &=  0.167 \left(\frac{r}{R}\right)^2 + 0.370 \left(\frac{r}{R}\right)^4 + 1.187 \left(\frac{r}{R}\right)^{6} + \mathcal{O}\left(\left(\frac{r}{R}\right)^{8}\right).
\end{align}
For $\alpha=4$, we have,
\begin{align}
\begin{split}
    I_{2,4}(r, R) &= \left(\frac{1}{12}+\frac{2}{3 \pi^{2}}\right) \left(\frac{r}{R}\right)^2 + \left(\frac{3}{16}+\frac{4}{3 \pi^{2}}+\frac{4}{3 \pi^{4}}\right) \left(\frac{r}{R}\right)^4 \\& + \left(\frac{689}{1152}+\frac{112}{27 \pi^{2}}+\frac{16}{3 \pi^{4}}+\frac{32}{9 \pi^{6}}\right)\left(\frac{r}{R}\right)^{6} + \mathcal{O}\left(\left(\frac{r}{R}\right)^{8}\right) \\
    &=  0.151 \left(\frac{r}{R}\right)^2 + 0.336 \left(\frac{r}{R}\right)^4 + 1.077 \left(\frac{r}{R}\right)^{6} + \mathcal{O}\left(\left(\frac{r}{R}\right)^{8}\right).
\end{split}
\end{align}

Lastly for $\alpha=\infty$,
\begin{align}
\begin{split}
I_{2,\infty}(r, R) &= \left(\frac{4 \ln\! \left(2\right)}{\pi^{2}}
-\frac{1}{6}\right) \left(\frac{r}{R}\right)^2 + \left(\frac{20 \ln\! \left(2\right)}{3 \pi^{2}}+\frac{16 \ln\! \left(2\right)}{\pi^{4}}+\frac{6 \zeta \! \left(3\right)}{\pi^{4}}-\frac{1}{3}-\frac{2}{3 \pi^{2}}\right) \left(\frac{r}{R}\right)^4 \\
& + \left(\frac{84 \ln\! \left(2\right)}{5 \pi^{2}}+\frac{112 \ln\! \left(2\right)}{3 \pi^{4}}+\frac{64 \ln\! \left(2\right)}{\pi^{6}}+\frac{28 \zeta \! \left(3\right)}{\pi^{4}}+\frac{120 \zeta \! \left(3\right)}{\pi^{6}}+\frac{90 \zeta \! \left(5\right)}{\pi^{6}}-\frac{3767}{3780}-\frac{12}{5 \pi^{2}}-\frac{8}{3 \pi^{4}}\right) \left(\frac{r}{R}\right)^6 \\&+ \mathcal{O}\left(\left(\frac{r}{R}\right)^8\right) \\&
= 0.114 \left(\frac{r}{R}\right)^2 + 0.255 \left(\frac{r}{R}\right)^4 + 0.817 \left(\frac{r}{R}\right)^6 + \mathcal{O}\left(\left(\frac{r}{R}\right)^8\right).
\end{split}
\end{align}
\subsubsection{Three Dimensions}
For arbitrary  $\alpha$, we have

\begin{align}
\begin{split}
    I_{3,\alpha}(r, R) &=   \frac{(\alpha + 1) \left(\alpha^2 + 1\right)}{15 \alpha^3} \left(\frac{r}{R}\right)^4  + \frac{16 (\alpha+1)(4\alpha^2-1)(4\alpha^2+5)}{945 \alpha^5} \left(\frac{r}{R}\right)^6 \\ 
    & +\frac{4 (\alpha + 1) \left(29 \alpha^6 + 29 \alpha^4 - 13 \alpha^2 + 7\right)}{105 a^7} \left(\frac{r}{R}\right)^8 + \mathcal{O}\left(\left(\frac{r}{R}\right)^{10}\right)
\end{split}
\end{align}
Each of these terms agree with \cite{chen_odd}.

\subsubsection{Four Dimensions}
For  $\alpha = 2$ , we have,
\begin{align}
    I_{4,2}(r, R) &= \frac{8}{9 \pi^2} \left(\frac{r}{R}\right)^6 + \frac{16}{3 \pi^2} \left(\frac{r}{R}\right)^8 + \frac{6016}{225 \pi^2} \left(\frac{r}{R}\right)^{10} + \mathcal{O}\left(\left(\frac{r}{R}\right)^{12}\right) \\
    &= 0.0901 \left(\frac{r}{R}\right)^6 + 0.540 \left(\frac{r}{R}\right)^8 + 2.709 \left(\frac{r}{R}\right)^{10} + \mathcal{O}\left(\left(\frac{r}{R}\right)^{12}\right).
\end{align}
For  $\alpha = 3$ , we have,
\begin{align}
    I_{4,3}(r, R) &= \frac{1225}{17496} \left(\frac{r}{R}\right)^6 + \frac{177625}{419904} \left(\frac{r}{R}\right)^8 + \frac{3016097}{1417176} \left(\frac{r}{R}\right)^{10} + \mathcal{O}\left(\left(\frac{r}{R}\right)^{12}\right) \\
    &= 0.0700 \left(\frac{r}{R}\right)^6 + 0.423 \left(\frac{r}{R}\right)^8 + 2.128 \left(\frac{r}{R}\right)^{10} + \mathcal{O}\left(\left(\frac{r}{R}\right)^{12}\right).
\end{align}
For $\alpha=4$, we have,
\begin{align}
    I_{4,4}(r, R) &= \left(\frac{25}{768}+\frac{8}{27 \pi^{2}}\right) \left(\frac{r}{R}\right)^6 + \left(\frac{1625}{8192}+\frac{16}{9 \pi^{2}}\right) \left(\frac{r}{R}\right)^8 + \left(\frac{2053}{2048}+\frac{6016}{675 \pi^{2}}\right)\left(\frac{r}{R}\right)^{10} + \mathcal{O}\left(\left(\frac{r}{R}\right)^{12}\right) \\
    &= 0.0626 \left(\frac{r}{R}\right)^6 +0.378 \left(\frac{r}{R}\right)^8 + 1.905 \left(\frac{r}{R}\right)^{10} + \mathcal{O}\left(\left(\frac{r}{R}\right)^{12}\right).
\end{align}
Lastly for $\alpha=\infty$,
\begin{align}
    \begin{split}
I_{4,\infty}(r, R) &= \left(\frac{3 \zeta \! \left(3\right)}{\pi^{4}}+\frac{75 \zeta \! \left(5\right)}{2 \pi^{6}}+\frac{\ln\! \left(2\right)}{3 \pi^{2}}-\frac{407}{7560}\right) \left(\frac{r}{R}\right)^6  \\&+\left(\frac{87 \zeta \! \left(3\right)}{4 \pi^{4}}+\frac{975 \zeta \! \left(5\right)}{4 \pi^{6}}+\frac{5 \ln\! \left(2\right)}{2 \pi^{2}}-\frac{6615 \zeta \! \left(7\right)}{8 \pi^{8}}-\frac{20219}{60480}\right) \left(\frac{r}{R}\right)^8 \\&
+ \left(\frac{594 \zeta \! \left(3\right)}{5 \pi^{4}}+\frac{1220 \zeta \! \left(5\right)}{\pi^{6}}+\frac{14 \ln\! \left(2\right)}{\pi^{2}}-\frac{7497 \zeta \! \left(7\right)}{\pi^{8}}+\frac{16065 \zeta \! \left(9\right)}{\pi^{10}}-\frac{2128769}{1247400}\right) \left(\frac{r}{R}\right)^{10} + \mathcal{O}\left(\left(\frac{r}{R}\right)^{12}\right) \\&
= 0.047 \left(\frac{r}{R}\right)^6 + 0.285 \left(\frac{r}{R}\right)^8 + 1.434 \left(\frac{r}{R}\right)^{10} + \mathcal{O}\left(\left(\frac{r}{R}\right)^{12}\right).
    \end{split}
\end{align}
\subsubsection{Five Dimensions}
For arbitrary $\alpha$, we have 
\begin{align}
\begin{split}
       I_{5,\alpha}(r,R)&=\frac{(\alpha + 1) \left(103 \alpha^6 + 103 \alpha^4 + 61 \alpha^2 + 21\right) }{2835 \alpha^7}\left(\frac{r}{R}\right)^{8}\\&+\quad\frac{32 (\alpha + 1)(4\alpha^2-1) \left(356 \alpha^6 + 445 \alpha^4 + 294 \alpha^2 + 105\right) }{155925 \alpha^9}\left(\frac{r}{R}\right)^{10} \\&+\quad\frac{4 (\alpha + 1) \left(535553 \alpha^{10} + 535553 \alpha^8 + 247265 \alpha^6 + 12745 \alpha^4 - 50318 \alpha^2 + 15202\right) }{1216215 \alpha^{11}}\left(\frac{r}{R}\right)^{12} \\& +\mathcal{O}\left(\left(\frac{r}{R}\right)^{14}\right).
\end{split}
\end{align}
Each of these terms agree with \cite{chen_odd}.
\subsubsection{Six Dimensions}
For $\alpha=2$, we have

\begin{align}
    I_{6,2}(r,R) &= \frac{128}{225 \pi^2} \left(\frac{r}{R}\right)^{10} + \frac{256}{45 \pi^2} \left(\frac{r}{R}\right)^{12} + \frac{146432}{3675 \pi^2} \left(\frac{r}{R}\right)^{14} + \mathcal{O}\left(\left(\frac{r}{R}\right)^{16}\right) \\
    &= 0.0576 \left(\frac{r}{R}\right)^{10} + 0.576 \left(\frac{r}{R}\right)^{12} + 4.037 \left(\frac{r}{R}\right)^{14} + \mathcal{O}\left(\left(\frac{r}{R}\right)^{16}\right).
\end{align}
For $\alpha=3$, we have
\begin{align}
    I_{6,3}(r,R) &= \frac{1002001}{22674816} \left(\frac{r}{R}\right)^{10} + \frac{1628251625}{3673320192} \left(\frac{r}{R}\right)^{12} + \frac{1428505793}{459165024} \left(\frac{r}{R}\right)^{14} + \mathcal{O}\left(\left(\frac{r}{R}\right)^{16}\right) \\
    &= 0.0442 \left(\frac{r}{R}\right)^{10} + 0.443 \left(\frac{r}{R}\right)^{12} + 3.111 \left(\frac{r}{R}\right)^{14} + \mathcal{O}\left(\left(\frac{r}{R}\right)^{16}\right).
\end{align}
For, $\alpha=4$ the expansion is
\begin{align}
\begin{split}
    I_{6,4}(r,R) &= \left(\frac{1323}{65536}+\frac{128}{675 \pi^{2}}\right) \left(\frac{r}{R}\right)^{10} +\left(\frac{106575}{524288}+\frac{256}{135 \pi^{2}}\right) \left(\frac{r}{R}\right)^{12} + \left(\frac{749943}{524288}+\frac{146432}{11025 \pi^{2}}\right) \left(\frac{r}{R}\right)^{14} \\&+\quad \mathcal{O}\left(\left(\frac{r}{R}\right)^{16}\right) \\
    &=  0.0394 \left(\frac{r}{R}\right)^{10} + 0.395 \left(\frac{r}{R}\right)^{12} + 2.776 \left(\frac{r}{R}\right)^{14} + \mathcal{O}\left(\left(\frac{r}{R}\right)^{16}\right).
\end{split}
\end{align}
Lastly for $\alpha=\infty$, 
\begin{align}
\begin{split}
    I_{6,\infty}(r,R) &=   \left(\frac{9 \zeta \! \left(3\right)}{8 \pi^{4}}+\frac{295 \zeta \! \left(5\right)}{16 \pi^{6}}+\frac{2205 \zeta \! \left(7\right)}{16 \pi^{8}}+\frac{16065 \zeta \! \left(9\right)}{32 \pi^{10}}+\frac{9 \ln\! \left(2\right)}{80 \pi^{2}}-\frac{128123}{3991680}\right)\left(\frac{r}{R}\right)^{10}
    +\biggr(\frac{2073 \zeta \! \left(3\right)}{160 \pi^{4}}\\&+\quad\frac{9875 \zeta \! \left(5\right)}{48 \pi^{6}}+\frac{42777 \zeta \! \left(7\right)}{32 \pi^{8}}+\frac{80325 \zeta \! \left(9\right)}{32 \pi^{10}}+\frac{21 \ln\! \left(2\right)}{16 \pi^{2}}-\frac{1181565 \zeta \! \left(11\right)}{64 \pi^{12}}-\frac{507316223}{1556755200}\biggr)\left(\frac{r}{R}\right)^{12}\\
    &+\biggr(\frac{13851 \zeta \! \left(3\right)}{140 \pi^{4}}+\frac{171531 \zeta \! \left(5\right)}{112 \pi^{6}}+\frac{8757 \zeta \! \left(7\right)}{\pi^{8}}+\frac{42687 \zeta \! \left(9\right)}{16 \pi^{10}}+\frac{81 \ln\! \left(2\right)}{8 \pi^{2}}-\frac{1519155 \zeta \! \left(11\right)}{8 \pi^{12}}\\&+\quad\frac{15810795 \zeta \! \left(13\right)}{32 \pi^{14}}-\frac{174511661}{75675600}\biggr)\left(\frac{r}{R}\right)^{14}\\
    & = 0.0296 \left(\frac{r}{R}\right)^{10} + 0.297 \left(\frac{r}{R}\right)^{12} + 2.086 \left(\frac{r}{R}\right)^{14} + \mathcal{O}\left(\left(\frac{r}{R}\right)^{16}\right).
\end{split}
\end{align}

\subsection{The Expansions for Rényi Entropy at Finite Temperature}\label{sec:RenyEntull}
\subsubsection{Three Dimensions}
For arbitrary $\alpha $, we have
\begin{align}
\begin{split}
    \Delta S_{3,\alpha}(r, T) &=   \frac{\pi^{2} \left(\alpha+1\right) }{18 \alpha} (rT)^2 
 + \frac{\pi^{4} \left(\alpha+1\right) \left(\alpha^{2}+1\right) }{1350 \alpha^{3}} (rT)^4 
    +\frac{4 \left(\alpha+1\right) \pi^{6} \left(\alpha^{4}+\alpha^{2}+\frac{2}{5}\right) }{5103 \alpha^{5}} (rT)^6
    \\ & -\frac{2 \left(\alpha+1\right) \pi^{8} \left(325 \alpha^{6}+325 \alpha^{4}+31 \alpha^{2}-69\right) }{1913625 \alpha^{7}} (rT)^8 \\
    &+\frac{16 \left(\alpha+1\right) \pi^{10} \left(521 \alpha^{8}+521 \alpha^{6}-111 \alpha^{4}-221 \alpha^{2}+10\right) }{63149625 \alpha^{9}}(rT)^{10}
+ \mathcal{O}\left((rT)^{12}\right)
\end{split}
\end{align}

\subsubsection{Four Dimensions}
For $\alpha=2$,
\begin{align}
    \begin{split}
        \Delta S_{4,2}(r,T) &= \frac{4 \zeta \! \left(3\right)}{3 \pi} (rT)^3 + \frac{32 \zeta \! \left(5\right)}{15 \pi} (rT)^5 - \frac{16 \zeta \! \left(3\right)^{2}}{9 \pi^{2}} (rT)^6 - \frac{64 \zeta \! \left(7\right)}{35 \pi} (rT)^7 + \frac{128 \zeta \! \left(3\right) \zeta \! \left(5\right)}{15 \pi^{2}} (rT)^8 + \mathcal{O}\left((rT)^9\right)\\
        &= 0.510 (rT)^3 + 0.704 (rT)^5 - 0.260 (rT)^6 - 0.587 (rT)^7 + 1.078 (rT)^8+\mathcal{O}\left((rT)^{9}\right)
    \end{split}
\end{align}
For $\alpha=3$,
\begin{align}
    \begin{split}
        \Delta S_{4,3}(r,T) &= \frac{35 \sqrt{3} \, \zeta \! \left(3\right)}{162} (rT)^3 + \frac{224 \sqrt{3} \, \zeta \! \left(5\right)}{729} (rT)^5 - \frac{1225 \zeta \! \left(3\right)^{2}}{8748} (rT)^6 - \frac{1856 \sqrt{3} \, \zeta \! \left(7\right)}{6561} (rT)^7\\& + \frac{4165 \zeta \! \left(3\right) \zeta \! \left(5\right)}{5832} (rT)^8 + \mathcal{O}\left((rT)^9\right)\\
        &= 0.450 (rT)^3 + 0.552 (rT)^5 - 0.202 (rT)^6 - 0.494 (rT)^7 + 0.890 (rT)^8+\mathcal{O}\left((rT)^{9}\right)
    \end{split}
\end{align}
For $\alpha=4$,
\begin{align}
    \begin{split}
        \Delta S_{4,4}(r,T) &= \left(\frac{5 \zeta \! \left(3\right)}{24} + \frac{4 \zeta \! \left(3\right)}{9 \pi} \right)(rT)^3 + \left(\frac{\zeta \! \left(5\right)}{4} + \frac{32 \zeta \! \left(5\right)}{45 \pi}\right)(rT)^5 + \left(-\frac{25 \zeta \! \left(3\right)^{2}}{384} - \frac{16 \zeta \! \left(3\right)^{2}}{27 \pi^{2}} \right)(rT)^6 \\
        &\quad + \left(-\frac{\zeta \! \left(7\right)}{4} - \frac{64 \zeta \! \left(7\right)}{105 \pi}\right)(rT)^7 + \left(\frac{365 \zeta \! \left(3\right) \zeta \! \left(5\right)}{1024} + \frac{128 \zeta \! \left(3\right) \zeta \! \left(5\right)}{45 \pi^{2}}\right)(rT)^8 + \mathcal{O}\left((rT)^9\right)\\
        &= 0.420 (rT)^3 + 0.494 (rT)^5 - 0.181 (rT)^6 - 0.448 (rT)^7 + 0.804 (rT)^8+\mathcal{O}\left((rT)^{9}\right)
    \end{split}
\end{align}
Lastly, $\alpha=\infty$,
\begin{align}
    \begin{split}
        \Delta S_{4,\infty}(r,T) &= \left(\frac{\ln\! \left(2\right) \zeta \! \left(3\right)}{\pi} + \frac{3 \zeta \! \left(3\right)^{2}}{2 \pi^{3}}\right)(rT)^3 + \left(\frac{4 \zeta \! \left(3\right) \zeta \! \left(5\right)}{\pi^{3}} + \frac{60 \zeta \! \left(5\right)^{2}}{\pi^{5}}\right)(rT)^5 \\
        &\quad + \left(\frac{407 \zeta \! \left(3\right)^{2}}{3780} - \frac{2 \ln\! \left(2\right) \zeta \! \left(3\right)^{2}}{3 \pi^{2}} - \frac{6 \zeta \! \left(3\right)^{3}}{\pi^{4}} - \frac{75 \zeta \! \left(3\right)^{2} \zeta \! \left(5\right)}{\pi^{6}}\right)(rT)^6 \\
        &\quad + \left(-\frac{56 \zeta \! \left(3\right) \zeta \! \left(7\right)}{5 \pi^{3}} - \frac{120 \zeta \! \left(7\right) \zeta \! \left(5\right)}{\pi^{5}} + \frac{1512 \zeta \! \left(7\right)^{2}}{\pi^{7}}\right)(rT)^7 \\
        &\quad + \biggr(-\frac{1223 \zeta \! \left(3\right) \zeta \! \left(5\right)}{1800} + \frac{52 \ln\! \left(2\right) \zeta \! \left(3\right) \zeta \! \left(5\right)}{5 \pi^{2}} + \frac{414 \zeta \! \left(3\right)^{2} \zeta \! \left(5\right)}{5 \pi^{4}} \\
        &\quad + \frac{630 \zeta \! \left(3\right) \zeta \! \left(5\right)^{2}}{\pi^{6}} - \frac{11907 \zeta \! \left(3\right) \zeta \! \left(7\right) \zeta \! \left(5\right)}{\pi^{8}}\biggr)(rT)^8 + \mathcal{O}\left((rT)^9\right)\\
        &= 0.335 (rT)^3 + 0.372 (rT)^5 - 0.136 (rT)^6 - 0.339 (rT)^7 + 0.607 (rT)^8+\mathcal{O}\left((rT)^9\right)
    \end{split}
\end{align}

\subsubsection{Five Dimensions}
For arbitrary $\alpha$, we have
\begin{align}
    \begin{split}
        \Delta S_{5,\alpha}(r,T) &= \frac{\pi^{4} \left(\alpha+1\right) \left(11 \alpha^{2}+1\right) }{4050 \alpha^{3}}(rT)^{4} + \frac{62 \left(\alpha^{4} + \alpha^{2} + \frac{10}{31}\right) \left(a + 1\right) \pi^{6} }{178605 \alpha^{5}}(rT)^{6} \\
        &\quad - \frac{\left(481 \alpha^{6} + 481 \alpha^{4} - 317\alpha^{2} - 357\right) \left(\alpha + 1\right) \pi^{8} }{11481750 \alpha^{7}}(rT)^{8} \\
        &\quad + \frac{2356 \left(\alpha^{4} - 1\right) \left(\alpha + 1\right) \left(\alpha^{4} + \alpha^{2} - \frac{170}{589}\right)  \pi^{10} }{189448875 \alpha^{9}}(rT)^{10} \\
        &\quad - \frac{4 \left(\alpha+1\right) \pi^{12}}{27152760009375 \alpha^{11}}(32883129 \alpha^{10} + 32883129 \alpha^{8} - 57623903 \alpha^{6} \\&\quad - 28582803 \alpha^{4} + 46609314 \alpha^{2} + 33550814)(rT)^{12}   \\
        &\quad + \mathcal{O}\left((rT)^{14}\right)
    \end{split}
\end{align}

\subsubsection{Six Dimensions}
For $\alpha =2$:
\begin{align}
    \begin{split}
        \Delta S_{6,2}(r,T) &= \left(\frac{16 \zeta \! \left(5\right)}{15 \pi}\right) (rT)^{5} + \left(\frac{64 \zeta \! \left(7\right)}{35 \pi}\right) (rT)^{7} - \left(\frac{512 \zeta \! \left(9\right)}{315 \pi}\right) (rT)^{9} \\
        &\quad - \left(\frac{256 \zeta \! \left(5\right)^{2}}{225 \pi^{2}}\right) (rT)^{10} + \left(\frac{2048 \zeta \! \left(11\right)}{693 \pi}\right) (rT)^{11} + \mathcal{O}\left((rT)^{12}\right)\\
        &= 0.352 (rT)^{5} + 0.587 (rT)^{7} - 0.518 (rT)^{9} + 0.124 (rT)^{10} + 0.941 (rT)^{11}+ \mathcal{O}\left((rT)^{12}\right)
    \end{split}
\end{align}
For $\alpha =3$, the expansion is
\begin{align}
    \begin{split}
        \Delta S_{6,3}(r,T) &= \left(\frac{1001 \sqrt{3} \zeta \! \left(5\right)}{5832}\right) (rT)^{5} + \left(\frac{572 \sqrt{3} \zeta \! \left(7\right)}{2187}\right) (rT)^{7} - \left(\frac{393536 \sqrt{3} \zeta \! \left(9\right)}{1594323}\right) (rT)^{9} \\
        &\quad - \left(\frac{1002001 \zeta \! \left(5\right)^{2}}{11337408}\right) (rT)^{10} + \left(\frac{6760000 \zeta \! \left(11\right) \sqrt{3}}{14348907}\right) (rT)^{11} + \mathcal{O}\left((rT)^{12}\right)\\
        &= 0.308 (rT)^{5} + 0.457 (rT)^{7} - 0.428 (rT)^{9} - 0.095 (rT)^{10} + 0.816 (rT)^{11}+ \mathcal{O}\left((rT)^{12}\right)
    \end{split}
\end{align}
For $\alpha =4$ it is
\begin{align}
    \begin{split}
        \Delta S_{6,4}(r,T) &= \left(\frac{21 \zeta \! \left(5\right)}{128} + \frac{16 \zeta \! \left(5\right)}{45 \pi}\right) (rT)^{5}+ \left(\frac{64 \zeta \! \left(7\right)}{105 \pi} + \frac{27 \zeta \! \left(7\right)}{128}\right) (rT)^{7} \\&\quad-  \left(\frac{41 \zeta \! \left(9\right)}{192} + \frac{512 \zeta \! \left(9\right)}{945 \pi}\right) (rT)^{9}- \left(\frac{256 \zeta \! \left(5\right)^{2}}{675 \pi^{2}} + \frac{1323 \zeta \! \left(5\right)^{2}}{32768}\right) (rT)^{10} \\
        &\quad +\left(\frac{55 \zeta \! \left(11\right)}{128} + \frac{2048 \zeta \! \left(11\right)}{2079 \pi}\right) (rT)^{11}   + \mathcal{O}\left((rT)^{12}\right)\\
        &= 0.287 (rT)^{5} + 0.408 (rT)^{7} - 0.387 (rT)^{9} - 0.085 (rT)^{10} + 0.744 (rT)^{11}+ \mathcal{O}\left((rT)^{12}\right)
    \end{split}
\end{align}
Lastly, for $\alpha =\infty$ we have:
\begin{align}
    \begin{split}
        \Delta S_{6,\infty}(r,T) &= \left(\frac{15 \zeta \! \left(5\right)^{2}}{8 \pi^{5}} + \frac{3 \zeta \! \left(5\right) \ln\! \left(2\right)}{4 \pi} + \frac{5 \zeta \! \left(3\right) \zeta \! \left(5\right)}{4 \pi^{3}}\right) (rT)^{5} \\
        &\quad + \left(\frac{45 \zeta \! \left(5\right) \zeta \! \left(7\right)}{\pi^{5}} + \frac{567 \zeta \! \left(7\right)^{2}}{4 \pi^{7}} + \frac{27 \zeta \! \left(3\right) \zeta \! \left(7\right)}{10 \pi^{3}}\right) (rT)^{7} \\
        &\quad + \left(-\frac{100 \zeta \! \left(5\right) \zeta \! \left(9\right)}{\pi^{5}} + \frac{504 \zeta \! \left(9\right) \zeta \! \left(7\right)}{\pi^{7}} + \frac{5100 \zeta \! \left(9\right)^{2}}{\pi^{9}} - \frac{264 \zeta \! \left(3\right) \zeta \! \left(9\right)}{35 \pi^{3}}\right) (rT)^{9} \\
        &\quad + \biggr(-\frac{295 \zeta \! \left(5\right)^{3}}{8 \pi^{6}} + \frac{128123 \zeta \! \left(5\right)^{2}}{1995840} - \frac{9 \zeta \! \left(5\right)^{2} \ln\! \left(2\right)}{40 \pi^{2}} - \frac{9 \zeta \! \left(3\right) \zeta \! \left(5\right)^{2}}{4 \pi^{4}} - \frac{2205 \zeta \! \left(5\right)^{2} \zeta \! \left(7\right)}{8 \pi^{8}}\\&\quad - \frac{16065 \zeta \! \left(5\right)^{2} \zeta \! \left(9\right)}{16 \pi^{10}}\biggr) (rT)^{10} 
        + \biggr(\frac{2240 \zeta \! \left(5\right) \zeta \! \left(11\right)}{9 \pi^{5}} - \frac{3696 \zeta \! \left(11\right) \zeta \! \left(7\right)}{\pi^{7}} - \frac{13600 \zeta \! \left(11\right) \zeta \! \left(9\right)}{\pi^{9}} \\&\quad + \frac{2504 \zeta \! \left(3\right) \zeta \! \left(11\right)}{105 \pi^{3}} + \frac{143220 \zeta \! \left(11\right)^{2}}{\pi^{11}}\biggr) (rT)^{11} + \mathcal{O}\left((rT)^{12}\right)\\
        &= 0.228 (rT)^{5} + 0.307 (rT)^{7} - 0.292 (rT)^{9} - 0.0636 (rT)^{10} + 0.564 (rT)^{11}+ \mathcal{O}\left((rT)^{12}\right)
    \end{split}
\end{align}

\bibliographystyle{JHEP.bst}

\bibliography{main.bib}

\end{document}